\documentclass[letterpaper,11pt]{article}
\usepackage{graphics,graphicx,epsfig,wrapfig}
\usepackage{tabularray}
\usepackage{multirow}
\usepackage[dvipsnames]{xcolor}
\usepackage{amsmath,verbatim,amssymb,color,lscape}
\usepackage{pdflscape}
\usepackage{kotex}
\usepackage{natbib}
\usepackage{float}
\usepackage{lastpage}
\usepackage{longtable}
\usepackage{authblk}
\usepackage{arydshln}
\usepackage{caption}
\usepackage{adjustbox}
\usepackage{subcaption}
\usepackage{bm}
\usepackage[ruled,vlined]{algorithm2e}
\usepackage[normalem]{ulem}
\usepackage{multirow}
\newcommand{\hide}[1]{}

\makeatletter

\usepackage[outerbars,color]{changebar}
\ifx\pdfoutput\undefined
\else\ifnum\pdfoutput>0
  \usepackage{pdfcolmk}
\fi\fi
\cbcolor{black}
\usepackage{caption}
\usepackage{subcaption}
\usepackage{url}

\usepackage[margin=1.0in]{geometry}
\newfont{\rmm}{cmr10 at 11pt}
\rmm

\title{Analysis of Log Data from an International Online Educational Assessment System: A Multi-state Survival Modeling Approach to Reaction Time between and across Action Sequence }

\author[1,2]{Jina Park}
\author[1,2]{Ick Hoon Jin}
\author[3]{Minjeong Jeon}
\affil[1]{Department of Applied Statistics, Yonsei University. South Korea.}
\affil[2]{Department of Statistics and Data Science, Yonsei University. South Korea.}
\affil[3]{School of Education and Information Studies, University of California, Los Angeles. USA.}
\date{}

\begin{document}
\maketitle

\begin{abstract}
With increasingly available computer-based or online assessments, researchers have shown keen interest in analyzing log data to improve our understanding of test takers' problem-solving processes. In this paper, we propose a multi-state survival model (MSM) to action sequence data from log files, focusing on modeling test takers' reaction times between actions, in order to investigate which factors and how they influence test takers' transition speed between actions. 
\textcolor{black}{
We specifically identify the key actions that differentiate correct and incorrect answers, compare transition probabilities between these groups, and analyze their distinct problem-solving patterns. Through simulation studies and sensitivity analyses, we evaluate the robustness of our proposed model. We demonstrate the proposed  approach using problem-solving items from the Programme for the International Assessment of Adult Competencies (PIAAC).
}
\end{abstract}
\newpage

\section{Introduction}\label{sec:intro}

Over the past decade, advances in technology have accelerated innovation in educational assessment, leading to the development of a growing number of computer-based problem-solving assessments that evaluate test takers' ability to solve complex problems in realistic environments. Examples of computer-based problem-solving assessments include the Program for the International Assessment of Adult Competencies (PIAAC), the Programme for International Student Assessment (PISA), and the National Assessment of Educational Progress (NAEP). A key feature of these assessments, compared with traditional paper-based assessments, is that user interactions with a testing system, such as clicking the button, dragging, dropping, and text input during assessments, are recorded in log files. This sequence of recorded user interactions, so-called log data or process data, is a valuable resource for various purposes, e.g., to explore and validate test takers' item-solving processes and strategies, identify key behaviors that determine the performance, and formulate real-time feedback to test takers \citep{Liu2018, Han2019, Jiao2021, He2021, Ulitzsch2021, Xiao2021}.

Analyzing log data using traditional statistical models, such as generalized linear models and item response theory models, is challenging due to the non-standard format, varying sequence lengths between participants, and high computational requirements, among others \citep{Tang2020c, Tang2021, Zhan2022, Xiao2023}. Researchers have proposed various methodologies to address these challenges in log data analysis, which could be classified into two types: 1) extraction of behavioral characteristics and 2) psychometric modeling of log data \citep{Han2022, Fu2023, Xiao2023}. We briefly review these two types of methods below. 

First, methods for extracting behavioral characteristics from log data fall into theory-based or data-driven approaches \citep{Yuan2019, Han2022, Fu2023}. Theory-based methods typically use expert-defined behavioral indicators, and thus different feature extraction rules are used for different problem-solving tests. For example, \citet{Greiff2015, Greiff2016} defined the optimal exploration strategy (e.g., vary-one-thing-at-at-time; VOTAT), time on task, and intervention frequency to examine their relationships with problem-solving test performance. Data-driven approaches, on the other hand,  employ data mining, machine learning, and other statistical methods to extract features from log data. For example, \citet{Tang2020b} used multidimensional scaling (MDS) to standardize varying lengths of log sequences. \citet{Tang2020c} employed a sequence-to-sequence autoencoder that encodes log sequences as numeric vectors. \citet{Zhu2016} and \citet{Vista2017} used network analysis to visualize log sequences and extract meaningful information from log data. \citet{Qiao2018} applied supervised learning, such as classification and regression trees (CART), gradient boosting random forests, and support vector machines (SVMs), and unsupervised learning, such as self-organizing map (SOMs) and K-means, to log data, evaluating consistency of the results across methods. \citet{He2019} utilized the longest common subsequence (LCS) method to define the optimal sequence from the log data. \citet{Xu2020} applied a latent topic model with a Markov structure, which extends the hierarchical Bayesian topic model with a hidden Markov structure, to obtain latent features of log data. 

Second, psychometric modeling of log data has typically focused on estimating test-takers' latent traits from log data. For example, \citet{Shu2017} developed a Markov item response theory model that combines Markov models with item response theory to identify latent characteristics of test takers and the tendency of each transition to occur. \citet{Han2022} applied mixture Rasch models to log data, specifically mixture partial credit models, to estimate latent features of students. \citet{Han2021} proposed a sequential response model (SRM) that combines a dynamic Bayesian network with a psychometric model to infer test takers' continuous latent abilities from log data.

Time information has been recognized as a critical element in the analysis of test-taking behavior, which offers useful information about the engagement and performance of the respondents \citep{Goldhammer2014, Scherer2015, Vrs2016, He2019b, Engelhardt2019}. Researchers have employed various methodologies to examine the relationship between time-related factors and test outcomes. For example, generalized linear mixed models have been used to study the effect of time on task in computer-based assessments of reading and problem-solving \citep{Goldhammer2014}, while two-level response time item response theory models have explored the connection between problem-solving time and ability by modeling dichotomous item responses and log-transformed reaction times jointly \citep{Scherer2015}. Studies have consistently found that an increase in time investment often correlates with higher test scores \citep{He2019b}. Other studies used time data to validate the interpretation of test scores for skills such as reading and reasoning \citep{Engelhardt2019}, and investigated the relationship between response time and action frequency, along with the combined effect of these factors on task performance \citep{Vrs2016}. 

Importantly, timestamps from log data can provide crucial insights into the efficiency and fluency of cognitive processing of respondents \citep{Xu2020, Wang2022}. These temporal markers are particularly valuable in multi-step or problem-solving tasks, where transition speeds between actions can indicate respondents' proficiency. Recent studies have leveraged timestamped log data using a variety of methodologies, including continuous-time dynamic choice models \citep{Chen2020}, latent topic models with Markov transitions \citep{Xu2020}, sequence mining techniques \citep{Ulitzsch2021}, and joint models of action sequences and times \citep{Fu2023}. These approaches have allowed researchers to estimate latent abilities and behavioral speeds \citep{Chen2020}, cluster learning trajectories \citep{Xu2020}, investigate the behavioral patterns of correct and incorrect groups \citep{Ulitzsch2021}, segment long processes into interpretable subprocesses \citep{Wang2022}, and identify different problem-solving strategies \citep{Zhang2023}.

Analysis of transition times between actions can offer a comprehensive view of respondent behavior in educational assessments. This approach can go beyond evaluating respondents' speed and further illuminate how respondents navigate tasks and employ cognitive strategies. By examining the temporal patterns of transitions in problem-solving tasks, we can also discern crucial differences in approach between correct and incorrect answer groups.

This paper introduces a novel approach to analyzing the impact of various factors on action transition speeds in log data, focusing on differences in transition patterns between correct and incorrect answer groups. We propose a multi-state survival model \citep[MSM]{Commenges1999, Hougaard1999, Andersen2002, Putter2006, MeiraMachado2008, Crowther2017} to overcome the limitations of traditional survival methods when dealing with non-standard log data formats. Our model examines progression through multiple states over time, particularly evaluating how key actions influence transition speeds. Rather than analyzing all possible actions, which would be conceptually and computationally inefficient given the large number and variety of actions in typical item solution processes, we concentrate on the effects of `key actions' at the start and end of transitions. These key actions are identified using $\chi^2$ statistical approach \citep{He2015} as significantly differentiating between correct and incorrect responses.

The rest of the paper is organized as follows. In Section 2, we start by describing the motivating data example. We explain how we extract key actions for the selected test items. 
\textcolor{black}{In Section 3, we provide an overview of MSM and its traditional applications, highlighting theoretical and practical justification of our novel MSM application to log data. }
In Section 4, we present the proposed MSM approach for time sequence data across actions and explain the Bayesian estimation approach for the proposed model. In Section 3, we describe the application of the proposed model to the motivating data example. \textcolor{black}{In Section 5 presents simulation studies and prior sensitivity analyses to evaluate model performance and robustness.} Finally, we conclude the paper with a summary and discussion in Section \textcolor{black}{6}.

\section{Motivating Examples} \label{sec:piaac}

\subsection{PIAAC Problem Solving Test}

The Organization for Economic Cooperation and Development (OECD) has implemented the Program for the International Assessment of Adult Competencies (PIAAC) for adults from over 40 countries since 2011 \citep{piaac2017}. PIAAC measures adult literacy, numeracy, and problem-solving skills in technology-rich environments (PSTRE) and examines how adults apply these skills in a variety of areas, including home, work, and community. 

We utilize the PSTRE assessment that focuses on `\emph{using digital technology, communication tools, and networks to acquire and evaluate information, communicate with others, and perform practical tasks}' \citep{OECD2011, OECD2012, OECD2016}. PSTRE evaluates individuals' problem-solving skills across various domains, including personal and professional domains, using computers. During PSTRE evaluation, user interactions, such as button clicks, links, drag, drop, copying, and pasting, are automatically logged into a separate log file with a timestamp. 

\begin{figure}[htb!]
\begin{subfigure}[b]{0.48\textwidth}
    \centering
    \includegraphics[width = \textwidth]{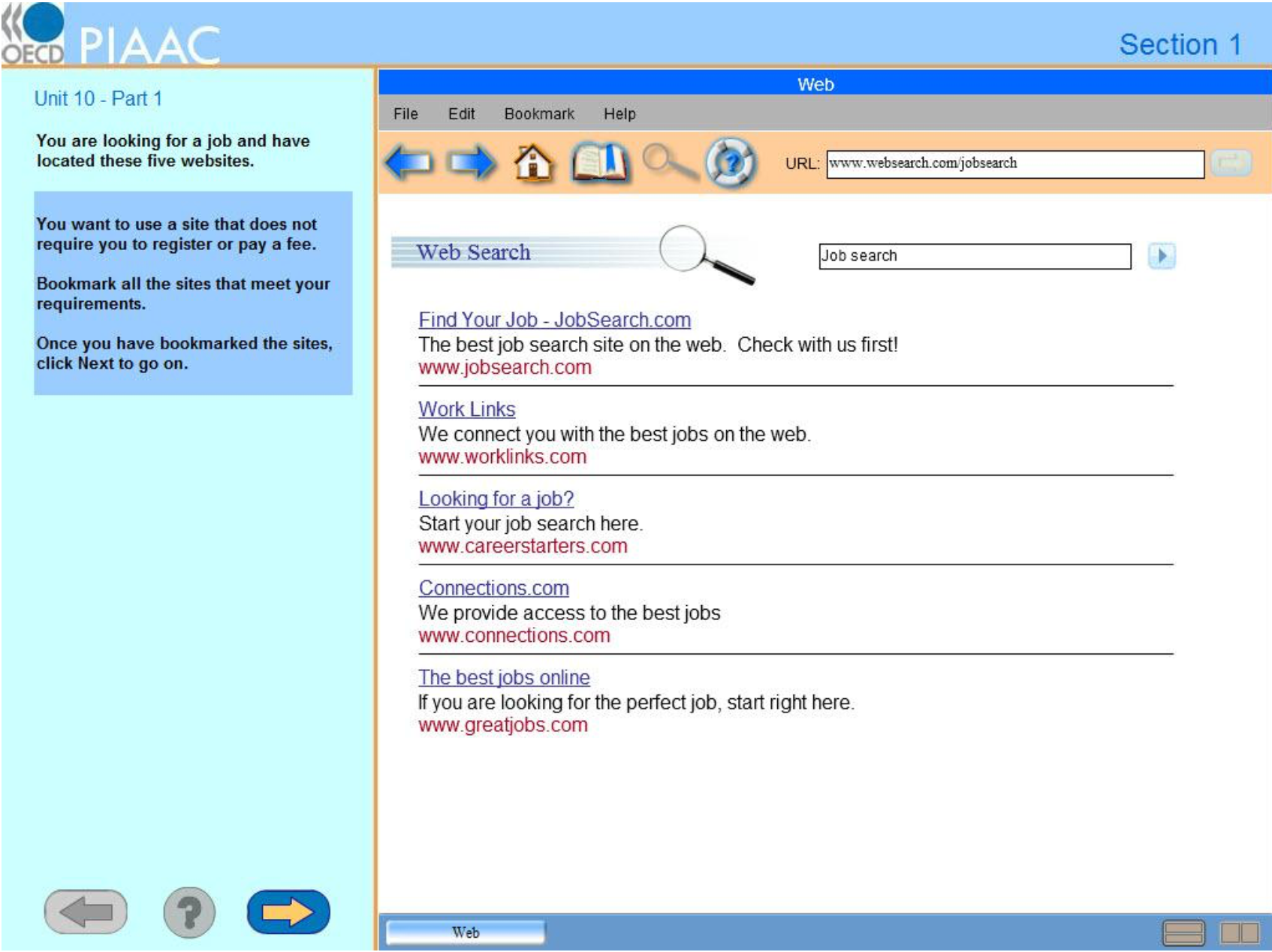}
\end{subfigure}  
\hfill
\begin{subfigure}[b]{0.48\textwidth}
    \centering
    \includegraphics[width = \textwidth]{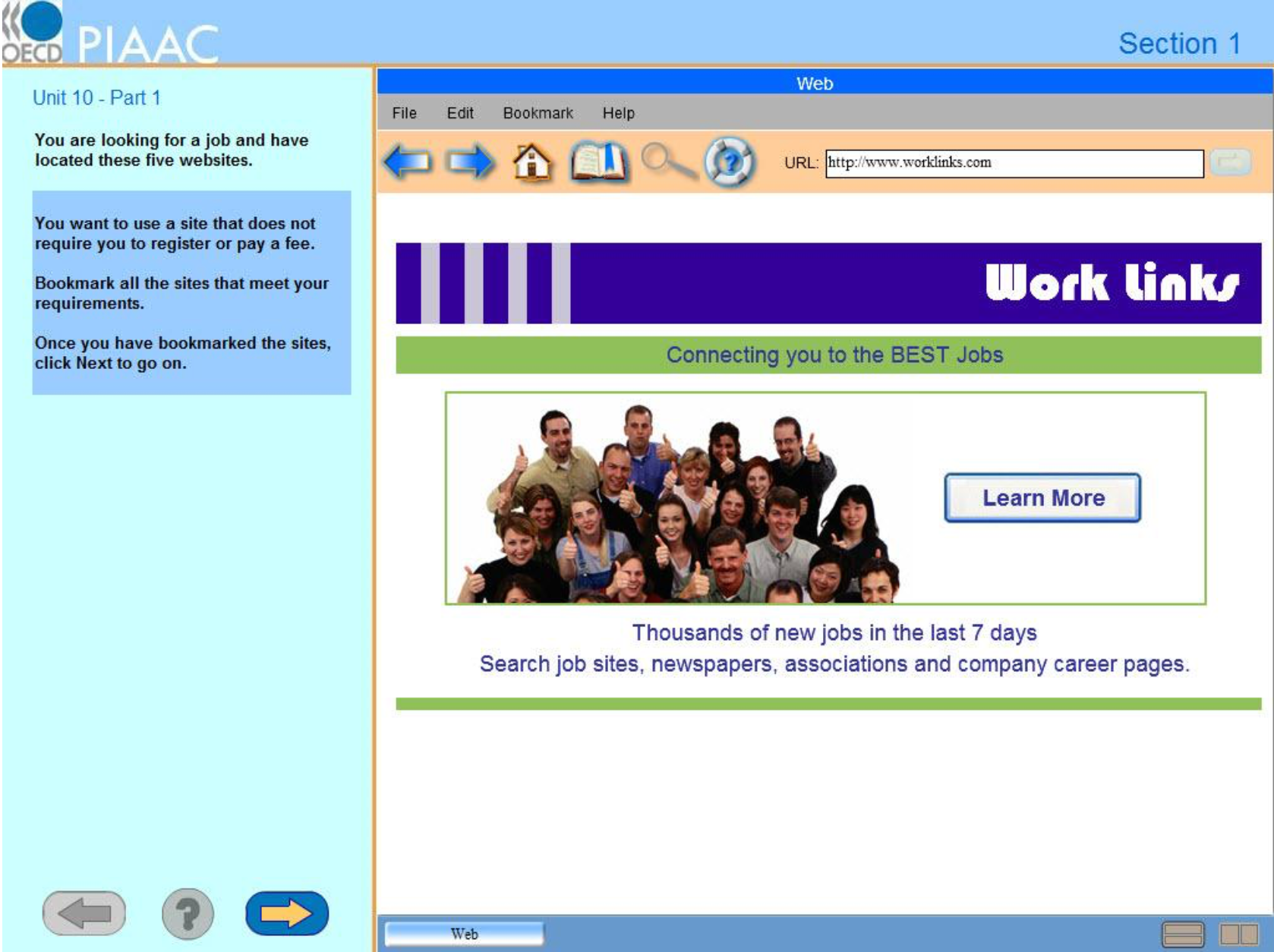}
\end{subfigure}
\caption{\label{tab:piaac-ex} 
A publicly available example of the PSTRE assessment of 2012 PIAAC about simulated job searches. Figure (a) and (b) are the list of job search sites and the page for the first link, respectively. 
}
\end{figure}

In addition, we use the public use file (PUF), which includes PIAAC participants' background information, including employment, income, health, education, and technology used in work and life. We selected four variables from the PUF and defined the `Eskill' variable by aggregating seven ICT-related variables from the PUF, which measure respondents' frequency of using various computer and Internet technologies for work. These seven variables related to ICT include the frequency of using email, work-related information, conducting transactions, spreadsheets, word processing, programming languages, and real-time discussions, all rated on a scale of 1-5. We normalize the `Eskill' variable, which initially ranges from 1 to 35, to prevent it from disproportionately influencing the analysis due to its larger range. Table \ref{tab:cov} provides detailed information on all selected variables, including descriptions, measurement scales, and descriptive statistics. Our final analysis includes 9,117 participants for the CD Tally test item and 12,557 participants for the Lamp Return test item, all of whom provided responses to all selected variables and the respective test items.

\begin{table}[htp!]
    \centering
    {\footnotesize %
        {
        \begin{tabular}{p{1.6cm}|p{4.2cm}|p{6.6cm}|p{2cm}}
            Covariate & Description & Value & Mean (SD) \\ \hline
            Gender & Gender & Male (1); Female (2) & 1.454 (0.497)\\
            Age & Age in 5-year intervals & 16-19 (1) - 60-65 (10) & 4.826 (2.314) \\
            Education & Highest schooling level & Less than high school (1) - Above high school (3) & 2.541 (0.604) \\
            IncPctRank & Income percentile rank & 6 categories from $<$ 10 (1) - 90+ (6) & 3.561 (1.611) \\
            Eskill & Internet/computer proficiency & Standardized the aggregated variable from 7 items measuring frequency of internet/computer use & 0.341 (0.865)\\
        \end{tabular}
        }
    }
    \caption{
        The description and mean (standard deviation) of the five selected covariates from the Public Use Files. The total numbers of participants for the CD Tally and Lamp Return test items are 9,117 and 12,557, respectively.
    }
    \label{tab:cov}
\end{table}

\subsection{CD Tally and Lamp Return}

The PSTRE assessment from 2012 PIAAC consists of 14 problem-solving items. These items are typically designed based on four specific environments: email, web browsing, word processing, and spreadsheet. Figure \ref{tab:piaac-ex} displays a publicly available example PSTRE test item in which participants engage in simulated job searches. In this item, participants are instructed to find a website that does not require registration or payment and then bookmark it. In order to solve this item successfully, participants need to navigate multiple website pages and bookmark the websites that do not require any registration or payment.

In this paper, we consider two problem-solving items, CD Tally (Item ID: U03A) and Lamp Return (Item ID: U23X) among the 14 PSTRE assessments from the 2012 PIAAC. Among PIAAC's four environmental designs (email, web browsing, word processing, and spreadsheet), CD Tally item is based on web and spreadsheet applications, and the Lamp Return item is based on email and web applications. The following is a detailed description of each item.

\paragraph*{CD Tally}

In the CD Tally item, test takers are asked to update the store's online inventory as requested by the manager. The CD Tally test item contains two pages: a website and a spreadsheet. The spreadsheet contains various details about the CDs, such as title, artist, genre, and release date. The goal is to count the number of CDs in the blues genre in the spreadsheet and enter them into the website. A total of 52 actions are identified for the CD Tally test item, which are detailed in the Section 1 of the Supplementary Material. {An example of log data for CD Tally is `wb (1.33) - ss (1.37) - ss\_file (1.86) - so (1.95) - so\_1\_3 (2.02) - so\_2\_asc (2.52) - so\_ok (2.57) - wb (2.67) - combobox (2.76)', where value in parentheses indicate the time (min) that the action occurred. This log shows the process of the participant's selecting sort options, sorting the spreadsheet, and entering the answers into a combobox. 

\paragraph*{Lamp Return}

The Lamp Return item assumes that the test taker receives a desk lamp in a different color from the one they ordered. The test taker is asked to request an exchange for a desk lamp in the correct color via the company's website. To accomplish this goal, the respondent clicks on the customer service page of the company's website and fills out a return form. The form requires an authorization number, which the participant receives via email. A total of 126 actions are identified for Lamp Return. which is more than twice the CD Tally case. Details of the actions for Lamp Return are described in Section 1 of the Supplementary Material.

Table \ref{tab:correct} lists the number and percentage of correct and incorrect answers to the two items summarized for each of 14 countries. Note that in the Lamp Return test, responses are scored on a scale from 0 to 3, where higher scores indicate higher accuracy. We dichotomized the answers with a score of 3 as correct and considered other answers as incorrect.

\begin{table}[htb!]
\centering
\resizebox{.82\columnwidth}{!}{%
        \begin{tabular}{lccc|ccc}
            & \multicolumn{3}{c|}{CD Tally} & \multicolumn{3}{c}{Lamp Return}\\
            Country & \# of correct & \# of incorrect & \% of correct & \# of correct & \# of incorrect & \% of correct \\ \hline
            Austria & 437 & 217 & 0.67 & 262 & 560 & 0.32 \\
            Belgium & 342 & 199 & 0.63 & 304 & 443 & 0.41 \\ 
            Germany & 476 & 248 & 0.66 & 318 & 656 & 0.33 \\ 
            Denmark & 491 & 361 & 0.58 & 470 & 739 & 0.39 \\
            Estonia & 315 & 197 & 0.62 & 303 & 411 & 0.42 \\ 
            Finland & 449 & 249 & 0.64 & 499 & 432 & 0.54 \\
            United Kingdom & 625 & 364 & 0.63 & 534 & 817 & 0.40 \\
            Ireland & 298 & 189 & 0.61 & 249 & 446 & 0.36 \\
            South Korea & 447 & 165 & 0.73 & 316 & 528 & 0.37 \\ 
            Netherlands & 425 & 285 & 0.60 & 457 & 559 & 0.45 \\ 
            Norway & 540 & 319 & 0.63 & 429 & 623 & 0.41 \\ 
            Poland & 331 & 196 & 0.63 & 341 & 487 & 0.41 \\ 
            Slovakia & 275 & 121 & 0.69 & 222 & 370 & 0.38 \\ 
            United States & 335 & 221 & 0.60 & 267 & 515 & 0.34 \\
        \end{tabular}
    }
\caption{
The numbers and percentages of participants who answered CD tally and Lamp Return correctly and incorrectly in the {2012} PIAAC data. 
}
\label{tab:correct}
\end{table}

\subsection{Extracting Key Actions}\label{sec:key}

As described above, the two selected items, CD Tally and Lamp Return,  involve a large number of actions: 52 and 126 actions, respectively, for the item solution. Recognizing that not all actions contribute equally to the item solution process, we identify `key actions' to evaluate their influence on transition speed between actions.

Researchers have applied subjective or objective methods to extract key actions from action sequence log data. Subjective methods select key actions based on personal knowledge \citep{Greiff2015, Greiff2016}. Objective methods apply feature selection approaches for key action extraction. For example, \citet{Han2019} applied a random forest algorithm to extract the most predictive characteristics that distinguished the correct group from the incorrect group. \citet{He2015} utilized the $\chi^2$ statistics method and the weighted log-likelihood ratio test (WLLR) approach based on natural language processing. 

Here we apply \citet{He2015}'s $\chi^2$ statistic approach to select key actions that differentiate correct answers from incorrect answers for the two selected items. The $\chi^2$ statistic approach is based on the following four steps:
\begin{enumerate}
\item Calculate the inverse sequence frequency (ISF) for action $i$, $\text{ISF}_i = \text{log}(E/\text{sf}_i)$ where $\text{sf}_i$ is the occurrence frequency of action $i$.

\item Calculate the term frequency (TF), $\text{tf}_{ij}$,  which indicates the frequency of action $i$ for individual $j$. 

\item Combine ISF and TF to calculate weights as follows:
    \begin{equation*}
    \text{weight}(i,j) = \begin{cases}
    [1 + \log (\text{tf}_{ij})] \cdot \text{ISF}_i & \text{if } \text{tf}_{ij} \le 1 \\
    0 & \text{if } \text{tf}_{ij} = 0
    \end{cases}
    \end{equation*}

\item Calculate the chi-square score for each action with weighted frequencies.
\end{enumerate}

In the fourth step, a chi-square score is calculated using a $2 \times 2$ contingency table (given in Table \ref{tab:chi}), which is the frequency of crossing the presence of each action or action with the correctness or incorrectness of the response. $n_i$ and $m_i$ in Table \ref{tab:chi} indicate the weighted frequency of action occurrences in the correct and incorrect groups, respectively, and len($C_1$) and len($C_2$) in Table \ref{tab:chi} denote the sum of weighted frequency of occurrence in the correct and incorrect groups, respectively. 
\begin{table}[ht]
    \centering
    \begin{tabular}{c|cc}
         & Correct ($C_1$) & Incorrect ($C_2$) \\ \hline
        Action $i$ & $n_i$ & $m_i$ \\
        Except action $i$ & len$(C_1) - n_i$ & len$(C_2) - m_i$
    \end{tabular}
    \caption{The $2 \times 2$ contingency for chi-square test of action $i$.}
    \label{tab:chi}
\end{table}

The $\chi^2$ statistic approach aims to assess the independence of occurrence and correctness. Under the null hypothesis of independence, the chi-square score is given as:
\[
\chi^{2}=\frac{E\left(O_{11} O_{22}-O_{12} O_{21}\right)^{2}}{\left(O_{11}+O_{12}\right)\left(O_{11}+O_{21}\right)\left(O_{12}+O_{22}\right)\left(O_{21}+O_{22}\right)},
\]
where $O_{ij}$ represents the cell in the $i$th row and $j$th column of Table \ref{tab:chi}. Chi-square scores indicate the discriminatory power of actions in classification. Consequently, we organized the $\chi^2$ scores for each action in descending order. Additionally, if the ratio $n_i / m_i$ exceeds $\text{len}(c_1) / \text{len}(c_2)$, action $i$ is deemed more representative of the correct answer group ($C_1$). As a result, actions with a high $\chi^2$ score are selected among actions that satisfy $n_i/m_i > \text{len}(c_1) / \text{len}(c_2)$ as key actions.
\begin{figure}[htb!]
    \centering
    \begin{subfigure}{.45\textwidth}
         \centering
         \includegraphics[width = .95\textwidth]{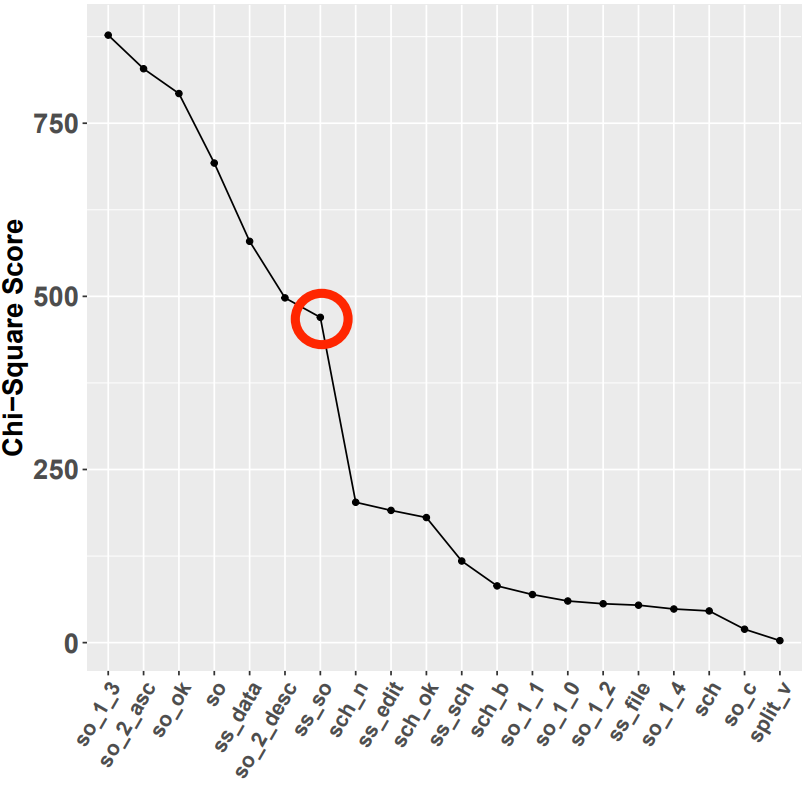}
         \caption{CD Tally}
         \label{fig:chi-line-u03a-1}
     \end{subfigure}
     \begin{subfigure}{.45\textwidth}
         \centering
          \includegraphics[width = .95\textwidth]{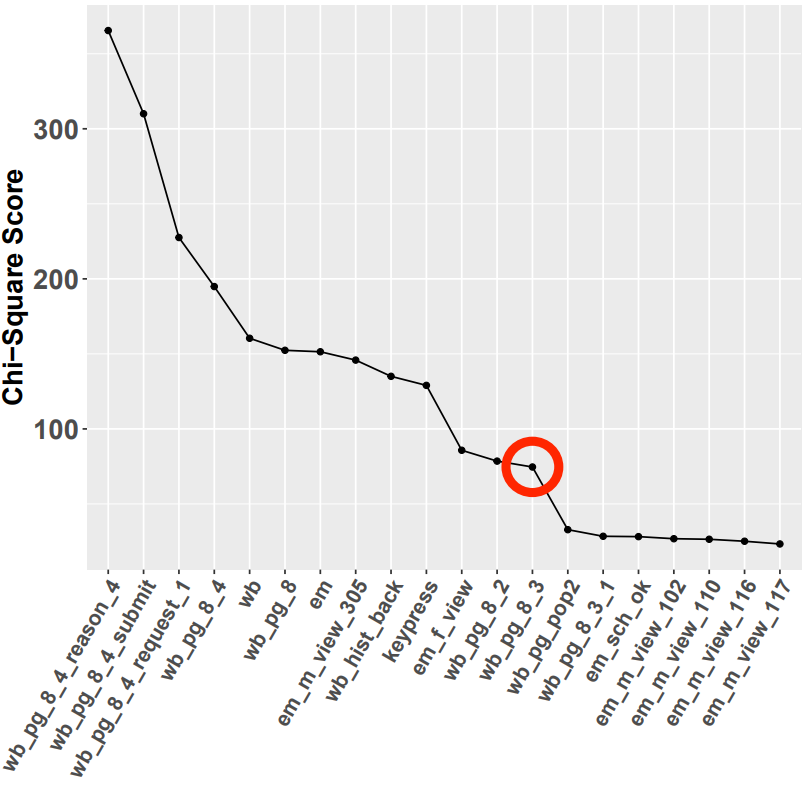}
         \caption{Lamp Return}
         \label{fig:chi-line-u03a-2}
     \end{subfigure}
    \caption{
        Line plot of chi-square scores in descending order for selecting key actions of (a) CD Tally and (b) Lamp Return test items. The red circle indicates the elbow point of the line plot. Actions with chi-square scores above this point are designated as key actions.
    }
    \label{fig:chi-line-u03a}
\end{figure}

To determine the cut-off point for selecting key actions, we use a line plot that visualizes the chi-square scores in descending order on the y-axis, where an elbow point is used as a threshold value. Figure \ref{fig:chi-line-u03a} shows line plots for selecting key actions of the two test items. The actions with a higher chi-square score than action `sch\_n' were selected as key actions.

\paragraph*{CD Tally: Key actions}

\begin{table}[ht]
\centering
\resizebox{\textwidth}{!}{%
    {
        \begin{tabular}{llccc}
            Action & Description & $\chi^2$ Score & Avg Freq & Avg Occur Time \\ \hline
            \textbf{so\_1\_3} & Sort by third column (Genre) & 877.19 & 0.37 & 1.97 \\ 
            \textbf{so\_2\_asc} & Sorts the spreadsheets in ascending order & 828.68 & 0.20 & 2.13 \\ 
            \textbf{so\_ok} & Click `Ok' after setting sorting options & 792.88 & 0.39 & 2.19\\ 
            \textbf{so} & Click the sort engine through the data menu on the spreadsheet page & 692.48 & 0.20 & 1.95 \\ 
            \textbf{ss\_data} & Click the data menu on the spreadsheet page & 579.46 & 0.15 & 2.04\\ 
            \textbf{so\_2\_desc} & Sorts the spreadsheets in descending order & 497.80 & 0.05 & 2.21\\ 
            \textbf{ss\_so} & Click the sort engine on the spreadsheet page  & 469.70 & 0.25 & 2.19\\ 
            sch\_n & Click the next button in the search engine & 202.66 & 0.08 & 2.74\\ 
            ss\_edit & Click the edit menu on the spreadsheet page & 190.94 & 0.09 & 2.24\\ 
            sch\_ok & Click "OK" after writing a search topic & 180.62 & 0.13 & 2.62\\ 
            ss\_sch & Click the search engine on the spreadsheet page & 117.84 & 0.09 & 2.20\\ 
            sch\_b & Click the back button in the search engine & 81.91 & 0.03 & 3.19 \\ 
        \end{tabular}
    }
}
\caption{
The top 12 actions based on the chi-square scores, along with the average occurrence frequency and average occurrence time (min) per person for the CD Tally test item. The 7 actions, marked in bold, are selected as key actions for this test item. 
}
\label{tab:chi-u03a}
\end{table}

Table \ref{tab:chi-u03a} lists the top 12 actions based on $\chi^2$ scores for the CD Talley log data. Figure \ref{fig:chi-line-u03a} is a line plot of $\chi^2$ scores in descending order. In Figure \ref{fig:chi-line-u03a}, `ss\_so' represents the elbow point of the line plot. Actions with a higher $\chi^2$ score than `ss\_so' - namely `so\_1\_3', `so\_2\_asc', `so\_ok', `so', `ss\_data', `so\_2\_desc', and `so\_so' - are selected as key actions. For the CD Tally test item, the actions related to sorting spreadsheets are chosen as the key actions.

\paragraph*{Lamp Return: Key actions}

Table \ref{tab:chi-lamp} and Figure \ref{fig:chi-line-u03a} show the top 15 actions based on chi-square scores and a line plot of the descending chi-square scores to identify key actions for the Lamp Return test item, respectively. Actions with higher chi-square scores than action `wb\_pg\_8\_3', which is the elbow point in Figure \ref{fig:chi-line-u03a}, are selected as key actions. The selected key actions for Lamp Return are marked in bold in Table \ref{tab:chi-lamp}. The top ranked actions are related to the customer service page, including actions such as selecting a reason for the return, requesting a return, and submitting a return form. The key actions in the lower ranking are related to obtaining an authorization number by email.

\begin{table}[ht]
\centering
\resizebox{\textwidth}{!}{%
    {
        \begin{tabular}{llccc}
            Action & Description & $\chi^2$ Statistic & Avg Freq & Avg Occur Time \\ 
            \hline
            \textbf{wb\_pg\_8\_4\_reason\_4} & Select the reason for return (Wrong item shipped)  & 365.48 & 0.54 & 1.79 \\ 
            \textbf{wb\_pg\_8\_4\_submit} & Submit the return form & 310.03 & 0.62 & 2.52 \\ 
            \textbf{wb\_pg\_8\_4\_request\_1} & Select a request for returned items (Exchange for the correct item) & 227.54 & 0.57 & 1.85 \\ 
            \textbf{wb\_pg\_8\_4} & Link to view return form & 194.87 & 0.69 & 1.65 \\ 
            \textbf{wb} & Switch to website page & 160.41 & 1.45 & 1.94 \\ 
            \textbf{wb\_pg\_8} & Link to Customer Service page & 152.39 & 1.39 & 1.25 \\ 
            \textbf{em} & Switch to Email page & 151.42 & 1.43 & 1.76 \\ 
            \textbf{em\_m\_view\_305} & View email 305 (Confirm authorization number) & 145.86 & 0.63 & 1.96 \\ 
            \textbf{wb\_hist\_back} & Go to the previous page & 135.05 & 1.83 & 1.79 \\ 
            \textbf{keypress} & press the keyboard & 129.02 & 1.27 & 2.32 \\ 
            \textbf{em\_f\_view} & View email folder & 85.76 & 0.42 & 1.94 \\ 
            \textbf{wb\_pg\_8\_2} & Link to view updated orders and shipping information &78.55 & 0.15 & 2.00 \\ 
            \textbf{wb\_pg\_8\_3} & Link to obtain authorization number before returning & 74.62 & 0.91 & 1.54 \\ 
            wb\_pg\_pop2 & Click the close button on pop-up system message 2 & 32.86 & 0.96 & 1.72 \\ 
            wb\_pg\_8\_3\_1  & Request authorization number on the Customer Service page & 28.45 & 0.99 & 1.66 \\ 
        \end{tabular}
    }
}
\caption{
The top 15 actions based on the chi-square scores, along with the average occurrence frequency and average occurrence time (min) per person for the Lamp Return test item.  The top 13 actions, marked in bold, are selected as key actions for this test item.
}
\label{tab:chi-lamp}
\end{table}

\section{Multi-state Survival Models}\label{sec:background}

\textcolor{black}{
In this paper, we adapt a multi-state survival model \citep[MSM;][]{Commenges1999, Hougaard1999, Andersen2002, Putter2006, MeiraMachado2008, Crowther2017} to examine the influence of covariates and key actions on transition speeds between actions in log data, while identifying distinct transition patterns between correct and incorrect response groups. This approach effectively addresses the challenge of varying transition times across individuals, allowing for a comprehensive analysis of individuals' item solution process through multiple states over time. In this section, we begin by providing a detailed overview of the MSM, highlighting the novelty of our application of the MSM to log data analysis, which seeks to shed light on the complex dynamics of test takers' problem-solving behaviors. 
In Section \ref{sec:model}, we detail the formulation and estimation of the proposed MSM for log data. 
}

\subsection{MSM: Background}

The MSM approach \citep{Commenges1999, Hougaard1999, Andersen2002, Putter2006, MeiraMachado2008, Crowther2017} is a sophisticated analytical tool designed to analyze longitudinal failure time data, by modeling the progression of individuals through various states or phases over time, such as the progression of the disease. The MSM approach offers a framework for investigating individual differences in trajectories across different states and the effects of covariates on the transition between two states \citep{Crowther2017}.

MSM can be grouped into different types, based on assumptions on the dependency of transition hazard rates on time and memory properties. For example, Markov models assume future states depend only on the current state (memoryless property), whereas semi-Markov models allow transition intensities to depend on the duration in the current state by relaxing the memoryless property. Non-Markov models are the most general type, allowing for transition intensities to be dependent on the entire process history. Additionally, MSMs can be further classified based on time homogeneity, where time-homogeneous models assume constant transition rates over time, while time-inhomogeneous models allow transition rates to vary over time. Time-homogeneous models, often assuming Markov properties, offer simplicity and are suitable for systems with constant transition probabilities. In contrast, time-inhomogeneous models offer more flexibility with time-varying transition probabilities. We will utilize time-homogeneous Markov models, as elaborated in the following section.

In MSM, nonparametric approaches offer additional flexibility by making no distributional assumptions \citep{deWreede2010, Manevski2022}, while parametric approaches bring simplicity to the analysis through specific distributional constraints \citep{Krol2015}. Semi-parametric approaches, such as the Cox proportional hazards model, make fewer assumptions than parametric methods by not specifying the form of the baseline hazard function.  Semi-parametric approaches, such as the Cox proportional hazards model, make fewer assumptions by not specifying a baseline hazard function form \citep{Kneib2008}. These models are commonly employed to examine the relationship between covariates and the hazard rate in time-to-event data, and we adopt these models in the current study. 

\textcolor{black}{The Cox proportional hazards model defines the transition hazard from state $i$ to state $j$ at time $t$ as}
\begin{equation*}
    \lambda_{ij}(t|X) = \lambda_{ij0}(t)\exp (\boldsymbol{X \beta_k })
\end{equation*}
where $\lambda_{ij0}(t)$ is the baseline hazard hazard function, ${\bf X}$ represents $n \times p$ matrix of covariates, and $\boldsymbol{\beta_k}$ is the $p \times 1$ vector of regression coefficients for event $k$, with $n$ referring to the number of observations and $p$ indicating the number of covariates. This formulation allows for modeling transition hazards without specifying a parametric form for the hazard function.

Both frequentist and Bayesian approaches can be employed to estimate the hazard function in multi-state transitions. Maximum likelihood estimation (MLE), a frequentist method, finds parameter values that maximize the likelihood function based on observed data. It offers unbiased and efficient estimates under large samples and correct model specifications, and is easily implemented using standard software like the {\tt msm} package in the R software. However, MLE can be unreliable with sparse data, may fail to converge in complex models, and often assumes time homogeneity and independence, which are not always valid, and struggles to handle unexplained individual heterogeneity \citep{Shen2017, MatsenaZingoni2021}. On the other hand, Bayesian estimation yields consistent estimates even with small samples by incorporating prior knowledge and effectively handles individual heterogeneity and nonlinear covariates. It offers comprehensive parameter estimate distributions and robustness in complex models, though it is computationally intensive and sensitive to prior choices \citep{Zingoni2020}. We will go with Bayesian estimation as elaborated in Section \ref{sec:estimation}

{\color{black}

\subsection{MSM: Traditional Applications}

MSMs have traditionally been employed in medical and epidemiological research to analyze transitions between distinct health states over time \citep{Hougaard1999, Andersen2002, Crowther2017}. Specifically, MSMs are applied in contexts such as cancer progression, stroke recovery, and chronic disease management, where state transitions occur infrequently, may be only partially observable, and occur over extended timeframes \citep{Grevel2024}.

\begin{figure}[htbp]
    \centering
    \includegraphics[width=0.5\linewidth]{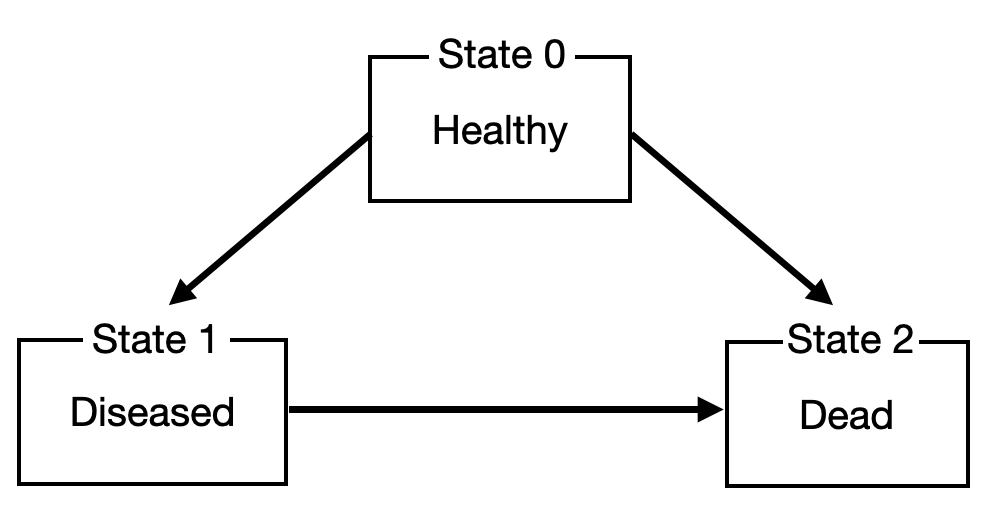}
    \caption{An illustration of the illness–death model}
    \label{fig:msm-ex}
\end{figure}

A well-known MSM application is the illness–death model, which captures an individual's progression through discrete health states. In its standard form, this model comprises three states: State 0 (healthy), State 1 (diseased) and State 2 (dead). Figure \ref{fig:msm-ex} depicts the typical structure of the illness-dealth model. Individuals may transition from state 0 to state 1 (onset of disease), from state 1 to state 2 (mortality after illness), or directly from State 0 to State 2 (death without prior disease). These transitions are generally considered irreversible, with the time intervals between states characterizing the temporal dynamics of disease progression.

Despite the widespread adoption of MSMs in medical and epidemiological research, their application have been limited to situations where there are a few clearly defined states and infrequent, unidirectional transitions. While these assumptions adequately capture long-term clinical processes, they constrain the utility of MSMs in domains characterized by rapid, high-resolution state changes. Consequently, despite the inherent flexibility of the model, researchers have yet to fully explore its use in settings characterized by complex behavioral dynamics and high-frequency event sequences.

}

{\color{black}

\subsection{MSM: Novel application to log data from online educational assessment systems}

The fundamental components of MSMs - discrete states, transition intensities, and covariate effects - can be effectively repurposed to analyze digital systems where entities transition over time. This is particularly relevant for digital log data from educational assessments, where users perform actions (e.g., clicking, dragging, copying) in rapid succession. Unlike traditional applications in clinical contexts, these datasets pose unique challenges: transitions occur at extremely high frequencies (often seconds apart), all state changes are fully observable (rather than partially observed), and the temporal patterns themselves carry meaningful information about cognitive processes. Although the semantic interpretation of states in digital environments differs from medical research, the underlying statistical framework of MSMs remains robust and adaptable. The contribution of the proposed model lies in extending and tailoring the MSMs to capture these distinctive features and the temporal dynamics of digital assessments.

This section explores the application of MSMs to the log data generated within online educational assessment systems. We first establish the theoretical foundation for employing MSMs in this context, with particular emphasis on their alignment with cognitive and educational frameworks. We then examine the unique insights these models can provide regarding respondents' behavioral patterns and instructional system design. 

\paragraph*{Theoretical Justifications}

Log data from educational assessments contain  time-stamped records of actions captured during test takers' problem-solving processes.  A key task is to infer the strategies and reasoning patterns that guide test takers' actions during the test. These actions are inherently sequential, as each decision depends on the outcomes of previous actions. Therefore, log data analysis is well-suited to modeling approaches that can capture temporal structures and transitions between behavioral states.

As noted in Section \ref{sec:intro}, various analytical approaches are available for log data, including latent topic modeling, sequence mining, and continuous-time dynamic modeling \citep[e.g.][]{Chen2020, Xu2020, Ulitzsch2021, Fu2023, Wang2022, Zhang2023}. These methods have successfully identified frequent behavioral patterns, clustered learning trajectories, predicted task performance, and jointly modeled response times and behavioral transitions. 

MSMs can provide a complementary analytical framework by explicitly modeling transition intensities between actions. Unlike methods primarily designed for clustering or prediction, MSMs uniquely capture both the timing and intensity of transitions while accounting for covariates that may influence transition probabilities. Furthermore, MSMs offer a powerful framework for modeling the dynamic progression of respondents through sequences of cognitive or behavioral states. This perspective is also consistent with cognitive theories such as information processing theory and cognitive load theory, which conceptualize cognition as an iterative process involving goal recognition, information structuring, action selection, and execution \citep{Dostal2015}. Importantly, MSMs effectively capture how individuals continuously monitor action outcomes and adapt their strategies through modeling of state transitions and their temporal characteristics.

According to cognitive load theory, excessive cognitive demands - particularly in complex tasks - can affect performance by overwhelming working memory capacity \citep{Paas2003}. However, directly measuring cognitive load from log data presents significant challenges due to the absence of explicit indicators \citep{Sweller1988}. MSMs represent a potentially promising alternative in this context, as they can capture both the temporal dynamics and structural patterns of transitions throughout problem-solving sequences—providing indirect yet meaningful indicators of cognitive effort at the individual level. For instance, rapid task completion with minimal planning may reflect disengagement or misunderstanding, whereas structured, stage-based progression may indicate deliberate strategy use. Transition parameters can be theoretically mapped to cognitive constructs: transition probabilities may index processing efficiency; the timing and sequence of transitions may align with executive functioning; and repetitive loops may signal confusion or strategy revision under cognitive load. This alignment enables interpretation of behavioral data within a psychologically grounded framework, beyond surface-level statistical patterns.

Therefore, modeling the transition times between specific actions allows for more fine-grained inferences about respondents' cognitive processing and problem-solving strategies. This framework also supports comparisons between correct and incorrect respondents, which can reveal differences in strategy patterns, processing efficiency, and decision pathways which cannot be captured by outcome measures alone. All together, MSMs offer a theoretically grounded and practically valuable approach for analyzing educational assessment log data. Hence, the proposed approach enables researchers to uncover subtle variations in respondents' cognitive processing and strategies that might otherwise remain undetected in conventional performance metrics.

\paragraph*{Research and Practical Contributions}

MSMs offer interpretive advantages that contribute meaningfully to both educational research and practice. From a research perspective, MSMs facilitate analysis of how students navigate complex tasks - beyond merely determining success, MSMs illuminate progression through each stage of the problem-solving process. For instance, in digital problem-solving items, MSMs can reveal whether high-performing students allocate more time to information review before taking action, or whether low-performing students systematically omit critical steps such as hypothesis generation or planning. This process-oriented analytical approach enables researchers to test and refine cognitive models for task performance and identify distinct behavioral patterns that differentiate effective from ineffective problem-solving strategies.

By modeling both the sequential structure and temporal dynamics of learner actions, MSMs transform behavioral log data into a theoretically grounded and diagnostically rich resource. This approach facilitates fine-grained comparisons between meaningful subgroups (e.g., correct vs. incorrect responders), uncovering cognitive strategies that remain obscured in traditional outcome-based analyses. For instance, our findings highlight critical transition points at which test takers  commonly experience difficulty, such as extended delays between task comprehension and strategy initiation, thereby identifying precise junctures where instructional scaffolding may be most impactful. Similarly, recurrent transitions between particular states (e.g., repeatedly returning to information sources without progressing) may signal conceptual confusion or indecisiveness.

These process-level insights can offer a foundation for the development of responsive educational systems that adapt in real time to learners’ behavioral indicators. Such applications are directly relevant to large-scale assessments and instructional platforms administered by organizations such as the OECD, which increasingly seek to integrate process data into evidence-based design. Specifically, MSM-informed diagnostics support:
\begin{itemize}
\item The design of targeted interventions addressing specific cognitive bottlenecks rather than generalized performance deficits,
\item The implementation of adaptive feedback mechanisms grounded in behavioral process markers rather than final outcomes,
\item The creation of interface designs that facilitate cognitively optimal task sequences, and
\item The deployment of analytics that clarify not only what learners struggle with, but why those struggles occur.
\end{itemize}

Ultimately, the proposed MSM approach moves beyond descriptive modeling to offer a principled, empirically grounded framework for advancing personalized learning, intelligent tutoring, and diagnostic assessment systems. MSMs thus represent a meaningful contribution to both theory-driven research and the practical optimization of educational assessment systems.

\section{Proposal: Multi-state Survival Model for Log Data}\label{sec:model}

\subsection{Model Formulation}

\textcolor{black}{
Log data consist of the sequential progression of actions for individuals. Our idea is to view individual actions as different states and apply MSM to the sequence of actions and their executed times. We can then model transition times between actions that respondents take while they are working on the problem-solving test items. 
}

To formulate the model, suppose $N$ is the number of respondents, $E$ is the total number of states (actions), and $E_i$ is the number of states that the respondent $i$ has gone through. Let $\boldsymbol{X_i} = \{x_{i,1}, x_{i,2}, \cdots, x_{i,P}\}$  is a vector of respondents' background characteristics and $\boldsymbol{A} = \{A_1, A_2, \cdots, A_K \}$ is a collection of key actions identified through a data-driven method as detailed in Section \ref{sec:key}. We adopted the time-homogeneous Markov assumption, which implies that transition rates remain constant over time and that future states depend solely on the current state. We then define the hazard function $\lambda_{m,l,i}$ for the transition from action $m$ to action $l$ for respondent $i$  as follows:
\begin{equation} 
    \label{equ:hazard}
    \lambda_{m,l,i} = \kappa_{c_i,m,l} \,\tau_i  \,\exp \Big\{ \sum_{p=1}^{P}\alpha_{p} x_{i,p}  +  \beta_{c_i, 1} I(m \in \boldsymbol{A}) +  \beta_{c_i, 2} I(l \in \boldsymbol{A})  \Big\}, 
\end{equation}
where $c_i$ is a binary indicator for correctness, with $c_i=0$ for incorrectness and $c_i=1$ for correctness, and $I(m \in \boldsymbol{A})$ and $I(l \in \boldsymbol{A})$ are indicator functions that equal 1 when start action $m$ and end action $l$ are key actions, respectively.

The model parameters, $\Theta = \{\boldsymbol{\kappa_{0}, \kappa_{1}, \tau, \alpha, \beta }\}$, are explained as follows: 
\begin{itemize}
    \item $\kappa_{c_i, m, l} > 0$ represents the baseline hazard for transitions between actions $m$ and $l$ for correct ($c_i=1$) and incorrect ($c_i=0$) groups. A larger $\kappa_{c_i,m,l}$ indicates a higher likelihood and faster rate of transitioning from action $m$ to action $l$, before accounting for the effects of covariates.
    \item $\tau_i> 0$ represents the overall speed of respondent $i$, with larger values indicating a tendency for respondents to transition between actions quickly. 
    \item $\boldsymbol{\alpha} = \{\alpha_1, \cdots, \alpha_P \}$ is a collection of the regression coefficients of respondents' background characteristics on \textcolor{black}{$\lambda_{m, l, i}$}. A greater $\alpha_{p}$ implies that individuals with a higher value for background ${x_{i,p}}$ have faster transition speeds compared to others. 
    \item $\beta_{c_i, 1}$ and $\beta_{c_i, 2}$ represent the effects when the start and end actions are key actions for group $c_i$, respectively. For group $c_i$, a larger $\beta_{c_i, 1}$ indicates a faster transition when the start action is the key action, while a larger $\beta_{c_i, 2}$ means a faster transition when the end action is the key action.
\end{itemize}

In our MSM for log data analysis, we estimate transition probabilities between actions under a time-homogeneous Markov assumption (the transition hazard rates between actions remain constant over time). This approach allows us to calculate the likelihood of a respondent moving from one action to another within an event sequence. Specifically, the transition probability $P_{m,l,i}$ for respondent $i$ moving from action $m$ to action $l$ is computed as
\begin{equation}\label{eq:prob}
    P_{m,l,i} = P(a_{i, j+1} = l, l \not=m | a_{i, j} = m) = \frac{\lambda_{m, l, i}}{\sum_{u=1}^{E}\lambda_{m, u, i}},
\end{equation}
where $a_{i, j}$ represent the respondent $i$-th action in their action sequence and $\lambda_{m, l, i}$ represents the transition hazard rate between behaviors $m$ and $l$. This transition probability effectively quantifies the relative likelihood of the next action being $l$ compared to all other possible actions.

\subsection{Estimation} \label{sec:estimation}

We estimate the model parameters of our proposed model using a fully Bayesian method via Markov chain Monte Carlo (MCMC). To define the likelihood function, we  denote $a_{i,j}$ represent the $j$-th action executed by respondent $i$, with $t_{i,j}$ indicating the action occurrence time. {Suppose $T_i$ represents the total time it took for individual $i$ to solve the item.}
We define $D_{m,l,i,j}$ as
\[
D_{m, l, i, j} = \left\{ \begin{array}{cl}
1 & a_{i,j-1} = m, ~ a_{i,j} = l,\\
0 & \mbox{otherwise}. 
\end{array} \right.
\]
That is, $D_{m, l, i, j}$ implies whether the transition from action $m$ to action $l$ is respondent $i$'s the $j$-th action  in his/her action sequence. In addition, we denote the risk set, $R_{m,i}(t) = \sum_{j=1}^{E_i}I\{a_{i,j-1}=m, t_{i,j-1} < t \leq t_{i,j}\}$, to be 1 when the respondent $i$ is in the state $m$ at time $t$.

In Bayesian estimation, the posterior distribution is derived from the product of the likelihood function and the prior distribution, enabling the integration of prior knowledge with observed data for accurate parameter estimation. The likelihood is similar to the traditional MSM model \citep{Hougaard1999, Hougaard2012}, with the primary distinction lying in the definition of the hazard function as presented in Equation \ref{equ:hazard}. The likelihood function of the proposed MSM for action transition from log data can be derived as follows:
\textcolor{black}{
\begin{equation*}\label{eq:likelihood}
\begin{split}
    L\Big({\cal Y}|\Theta\Big)&=
    \prod_{i=1}^N\prod_{m=1}^{E}\prod_{l=1, l\not=m}^{E} \left[ \Big(\prod_{j=1}^{E_i} \lambda_{m, l, i}^{D_{m, l, i, j}} \Big) \exp \Big( -\int_{0}^{T_i} R_{m, i}(t) \lambda_{m, l, i} dt \Big) \right], \\
    &=\prod_{i=1}^N\prod_{m=1}^{E}\prod_{l=1, l\not=m}^{E} \left[ \Big(\prod_{j=1}^{E_i} \lambda_{m, l, i}^{D_{m, l, i, j}} \Big) \exp \Big( -  \lambda_{m, l, i} \sum_{j=1}^{E_i}I(a_{i, j-1}=m)(t_{i, j} - t_{i, j-1})\Big) \right],
\end{split}
\end{equation*}
}
where ${\cal Y}$ represents the sequence of actions and their occurrence time, $\lambda_{m,l.i}$ is the hazard function as defined in Equation \ref{equ:hazard}, $T_i$ represents the time taken by respondent $i$ to solve the problem, and $\Theta = \{\boldsymbol{\kappa, \tau, \alpha, \beta }\}$ represents parameters of interest. 

The posterior distribution of $\Theta$ can then be written as follows: 
\begin{equation*}\label{eq:si_lsirm}
\begin{split}
    \pi\Big( \boldsymbol{\Theta} \mid {\cal Y}\Big) &\propto P\Big({\cal Y} \mid \boldsymbol{\Theta} \Big) \pi\Big(\boldsymbol\kappa\Big) 
    \pi\Big(\boldsymbol\tau\Big)
    \pi\Big(\boldsymbol\alpha\Big)
    \pi\Big(\boldsymbol\beta\Big)\\ 
    &= \prod_{i=1}^N\prod_{m=1}^{E}\prod_{l=1, l\not=m}^{E} \left[ \Big(\prod_{j=1}^{E_i} \lambda_{m,l,i}^{D_{m,l,i,j}} \Big) \exp \Big( -\int_{0}^{T_i} R_{m,i}(t) \lambda_{m,l,i} dt \Big) \right] \\
    & \times \pi(\boldsymbol\tau)
    \prod_{m=1}^{E} \prod_{l=1}^{E}\Big\{\pi(\kappa_{0, m, l}) 
    \pi(\kappa_{1, m, l})\Big\}
    \prod_{p=1}^{P} \pi(\alpha_p)
    \pi(\beta_{0, 1})
    \pi(\beta_{1, 1})
    \pi(\beta_{0, 2})
    \pi(\beta_{1, 2}), 
\end{split}
\end{equation*}
where the prior distributions for $\Theta$ are given as
\begin{equation*}
    \begin{split}
        \pi(\kappa_{\cdot m, l}) \sim \mbox{Gamma}&\big(a_{\kappa}, b_{\kappa}\big) , \quad 
        \pi(\tau_{i}) \sim \mbox{Gamma}\big(a_{\tau}, b_{\tau}\big) , \quad 
        \pi(\alpha_{p}) \sim \mbox{N}\big(0, \sigma_{\alpha} \big) , \\
        &\pi(\beta_{\cdot 1})  \sim \mbox{N}\big(0, \sigma_{\beta} \big), \quad 
        \pi(\beta_{\cdot 2}) \sim \mbox{N}\big(0, \sigma_{\beta} \big).\\
    \end{split}
\end{equation*}
We use $a_{\kappa} = b_{\kappa} = a_{\tau} = b_{\kappa}= 1.0$ and $\sigma_{\alpha} = \sigma_{\beta} = 2$ as the value of the prior distribution. Additional details of the MCMC sampling are in the Section 2 of the Supplementary Material} and the proposal distribution variances are adjusted to ensure moderate acceptance rates (approximately 0.2 - 0.5).

\section{Real Data Analysis}

We apply the proposed approach to analyzing the PIAAC problem-solving test items, CD tally and Lamp Return out of the log data from the 14 countries, described in Section \ref{sec:piaac}. The MCMC algorithm was iterated 300,000 times for each country and item, with the initial 100,000 iterations discarded as part of the burn-in process. Among the remaining 200,000 iterations, 20,000 samples were collected at 10-iteration intervals. Details of the jumping rules for the proposal distribution can be found in Section 2 of the Supplementary Material. 

\textcolor{black}{
To assess the convergence and reliability of parameter estimates, we executed five parallel MCMC chains and computed the Gelman-Rubin $\hat{R}$ statistics for all estimated parameters. Taking the CD Tally test item from the USA as an example, all $\hat{R}$ values remained below the commonly accepted threshold of 1.1, indicating satisfactory between-chain convergence. A histogram of the $\hat{R}$ values for all parameters in this case is presented in Figure 27 of Section 6 of the Supplementary Material. We also evaluated convergence across parallel MCMC chains using visual diagnostics. Convergence across parallel MCMC chains was further evaluated using visual diagnostics. Trace plots of MCMC samples (included in Section 6 of the Supplementary Material) demonstrated stable and consistent estimation across chains, confirming the reliability of the posterior estimates.
}

\begin{figure}[hbtp]
    \centering
    \includegraphics[width=\linewidth]{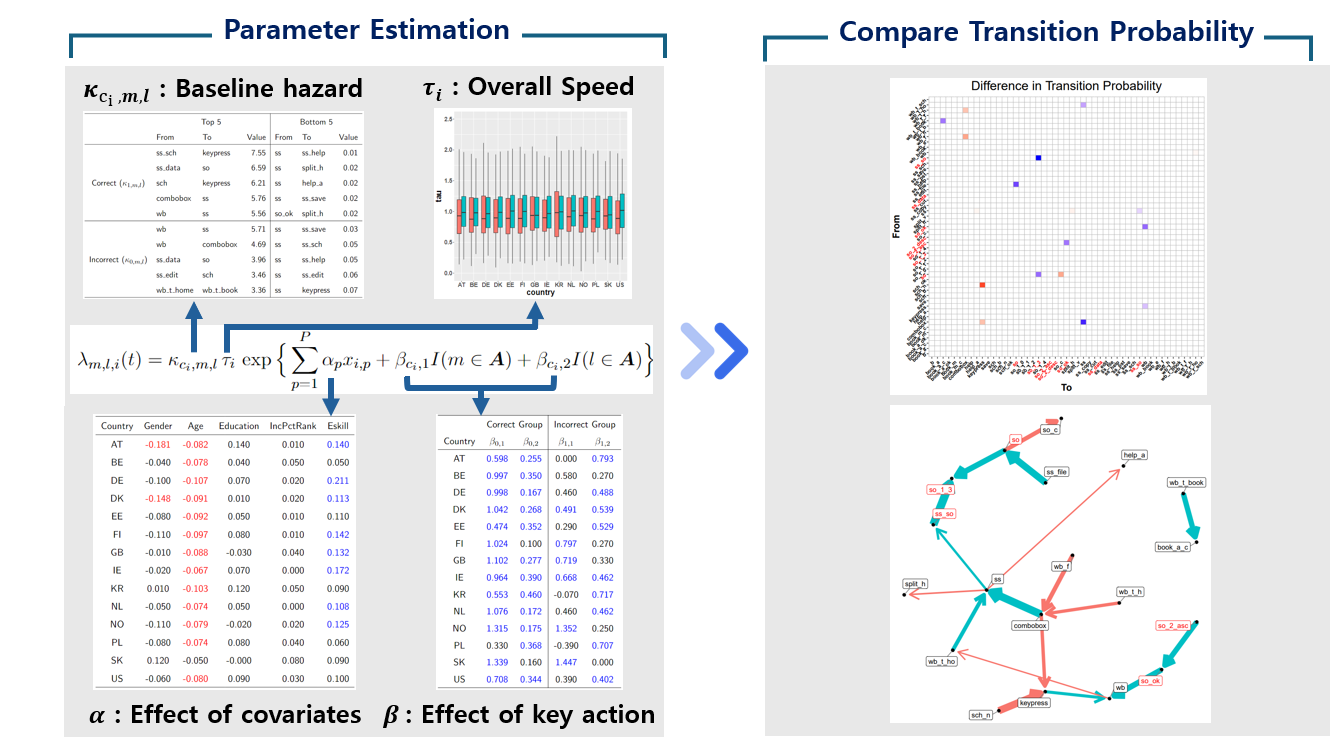}
    \caption{ The process of interpreting the result of the proposed model which consist of parameter estimation and comparing the transition probability between correct and incorrect group.}
    \label{fig:process}
\end{figure}

Figure \ref{fig:process} illustrates the process of interpreting results from the proposed method, including parameter estimation and comparison of transition probability between the correct and incorrect answer groups. Parameter estimation section summarizes estimated parameters including $\boldsymbol{\kappa}$, $\boldsymbol{\tau}$, $\boldsymbol{\alpha}$, and $\boldsymbol{\beta}$. For instance, $\boldsymbol{\kappa} \in \{\kappa_{c_i,m,l} \}$ denotes the baseline likelihood of transitioning from action $m$ to action $l$ for group $c_i$, excluding covariate effects. We present the top and bottom 5 $\kappa_{c_i,m,l}$ values for each group. Furthermore, $\boldsymbol{\alpha}$ represents covariate effects, while $\boldsymbol{\beta}$ captures the impact of start and end actions being key actions. These estimates are tabulated with color-coded significance levels: blue for positive effects, red for negative effects, and black for non-significant effects. For statistical inference, we use posterior means and 95\% Highest Posterior Density (HPD) intervals. We consider estimates with an HPD interval that includes 0 are considered statistically insignificant. 

Following parameter estimation, we compare correct and incorrect groups by calculating the transition probabilities between actions for each group. The estimated differences in transition probabilities are derived from all MCMC samples. Statistically significant differences are identified by checking whether the 95\% highest posterior density (HPD) interval includes zero. 

\textcolor{black}{We used both heatmap and network visualizations to highlight differences in transition probabilities.} 
In the heatmap, significant positive differences are represented in blue, negative differences in red, and non-significant differences in white. For instance, a blue color indicates that the transition probability for a specific transition is significantly higher in the correct answer group than in the incorrect answer group. Finally, we employ a network approach, representing action as nodes and transitions as directed edges. This network visualization makes it easy to identify notable differences in transition probabilities between actions. 
\textcolor{black}{While the heatmap provides a complete and compact overview of all transitions between actions, its interpretability decreases when the number of actions is large and the matrix becomes dense. The network plot, on the other hand, selectively displays only meaningful or strong transitions, improving readability in complex settings. Thus, the two visualizations serve complementary purposes, with the network becoming particularly valuable as the number of actions increases.}

Note that we fit the model to each country data separately, and thus, direct comparisons of the parameter estimates across countries are not desirable. We present the results from all countries with a goal in mind to identify any patterns in the results across the countries. 

\subsection{CD Tally}
\subsubsection{Parameter Estimation}
\begin{table}[htbp]
\centering
{\small {\color{black}
    \begin{tabular}{c|llr|llr}
        & \multicolumn{3}{c|}{Top 5} & \multicolumn{3}{c}{Bottom 5}\\
        & From & To & Value & From & To & Value \\ \hline
        \multirow{5}{*}{Incorrect ($\kappa_{0, m, l}$)} & 
        wb & ss & 5.86 & ss & ss\_save & 0.04 \\ 
        & wb & combobox & 4.82 & ss & ss\_help & 0.05 \\ 
        & ss\_edit & sch & 3.77 & ss & ss\_sch & 0.05 \\ 
        & ss\_data & so & 3.62 & ss & ss\_data & 0.05 \\ 
        & wb\_t\_home & wb\_t\_book & 3.41 & ss & ss\_so & 0.06 \\  \hline
        \multirow{5}{*}{Correct ($\kappa_{1, m, l}$)} 
        & ss\_sch & keypress & 8.09 & ss & ss\_help & 0.01 \\ 
        & sch & keypress & 6.49 & ss & help\_a & 0.02 \\ 
        & combobox & ss & 5.97 & ss & split\_h & 0.02 \\ 
        & wb & ss & 5.78 & so\_ok & ss\_so & 0.02 \\ 
        & ss\_data & so & 5.38 & ss & ss\_save & 0.02 \\ 
    \end{tabular}
}}
\caption{
The five highest and lowest $\kappa_{1, m, l}$ and $\kappa_{0, m, l}$ for correct and incorrect groups in the USA CD Tally test item.
}
\label{tab:kappa_ml}
\end{table}

\paragraph*{Parameter $\kappa_{c_i,m,l}$}
The parameters $\kappa_{c_i,m,l}$ denote the baseline hazard for transition from state $m$ to $l$ for correct ($c_i=1$) and incorrect ($c_i=0$) groups. The baseline hazard reflects the inherent probability of moving from action $m$ to $l$ when all covariates are zero. A higher value of $\kappa_{c_i,m,l}$ indicates a more frequent and rapid transition between these states, independent of covariate effects. 

Table \ref{tab:kappa_ml} presents the five highest and lowest values of $\kappa_{0,m,l}$ and $\kappa_{1,m,l}$ for incorrect and correct groups, respectively, using USA as an example (results for other countries are available in Section 4 of the Supplementary Material). For the correct group ($\kappa_{1,m,l}$), the fastest transition is from `ss\_sch' (clicking the search engine on a spreadsheet) to `keypress' (pressing the keyboard) with a value of \textcolor{black}{8.09}, indicating rapid transitions from search-related tasks to keyboard input. 
\textcolor{black}{
Other top five transitions include 'sch' (clicking the sort engine) to 'keypress' (pressing the keyboard), 'combobox' (interacting with a combobox on the webpage) to 'ss' (switching to the spreadsheet), 'wb' (switching to the webpage) to 'ss' (switching to the spreadsheet), and 'ss\_data' (viewing spreadsheet data) to 'so' (clicking the sort engine). Conversely, the bottom five transitions in the correct group predominantly originated from the spreadsheet state and involved auxiliary actions, including 'ss\_help' (clicking the help button for the spreadsheet), 'help\_a' (opening the general help page on how to answer), 'split\_h' (splitting the spreadsheet view horizontally), and 'ss\_save' (clicking the save button).
}

For the incorrect group ($\kappa_{0,m,l}$), the fastest transition is transition from `wb' (switching to the webpage) to `ss' (switching to the spreadsheet), with a value of \textcolor{black}{5.86}. Other rapid transitions include `wb' to `combobox' and \textcolor{black}{`ss\_edit' (click edit on the menu) to `sch'}. Similar to the correct group, the five slowest transitions for $\kappa_{0,m,l}$ predominantly involve spreadsheet action. This pattern suggests that transitions originating from `ss' actions occur at a slower pace in both correct and incorrect groups.

\paragraph*{Parameter $\tau$}
Parameter $\tau_i$ represents the overall action transition speed for the respondent $i$, with higher values indicating faster transitions. Figure \ref{fig:tau-u03a} shows boxplots of the estimated $\tau_i$ across countries (full country names are provided in the Section 3 of the Supplementary Material). 
\textcolor{black}{In Figure \ref{fig:tau-u03a}, the distributions are roughly symmetric and centered around 1 for all countries.}

\begin{figure}[htbp]
    \centering
    \includegraphics[width = .4\textwidth]{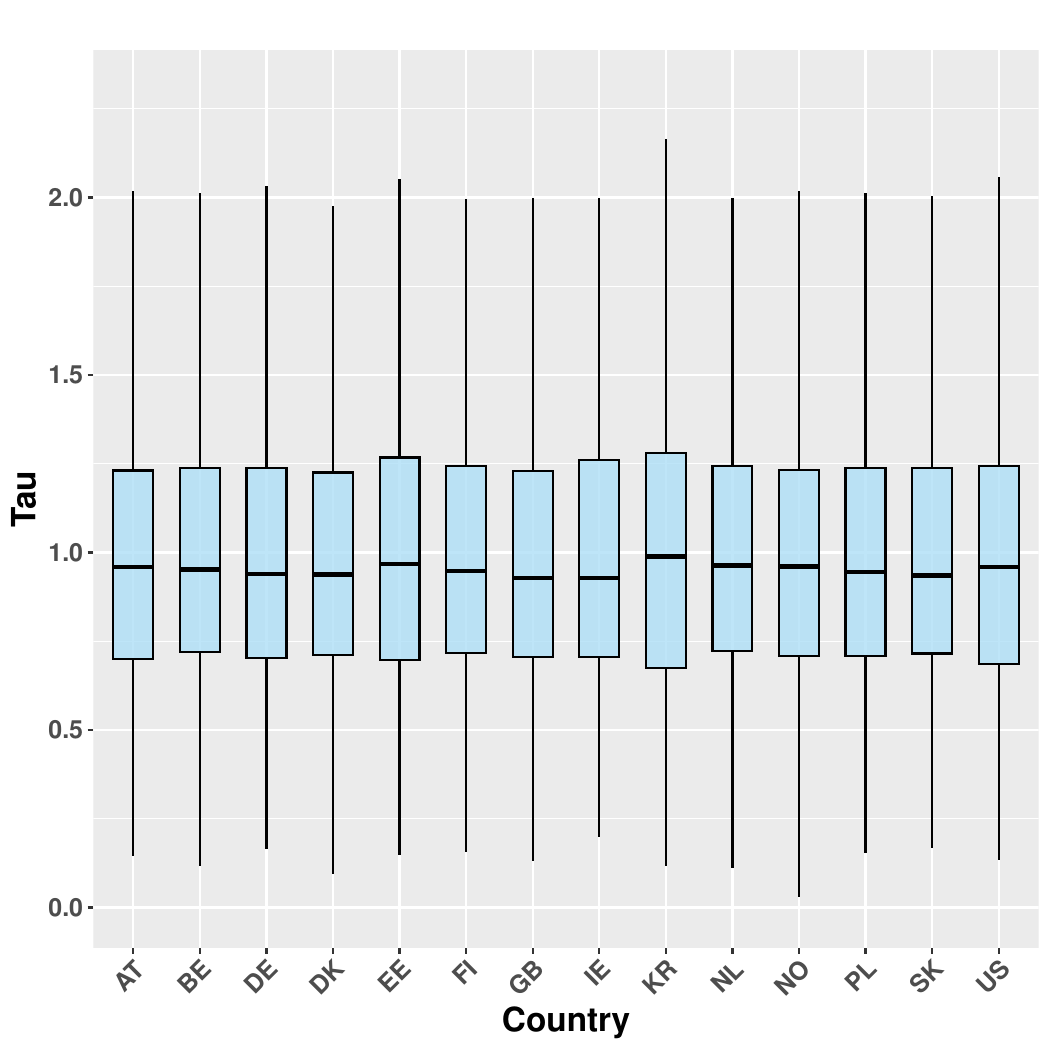}
    \caption{\textcolor{black}{The boxplots of posterior means of individual transition speed parameters ($\tau_i$) across the 14 countries for CD Tally test item.} Outliers omitted.}
    \label{fig:tau-u03a}
\end{figure}

\paragraph*{Parameters $\boldsymbol{\alpha}$}
The parameter $\boldsymbol{\alpha}$ quantifies the impact of various covariates on the action transition speed, with larger values indicating faster transitions for respondents with higher covariate values. The estimated $\boldsymbol{\alpha}$ is summarized in Table \ref{tab:beta-cov} with color-coded significance: blue for positive, red for negative, and black for non-significant. Across most countries, for CD Tally test item, age consistently demonstrates a negative impact on transition speed, suggesting older respondents solve problems more slowly. In contrast, `Eskill' positively influence speed, with more skilled individuals transitioning between actions more rapidly. Gender, education, and income percentile rank, however, generally show non-significant effects across countries.

\begin{table}[htbp]
\centering
{\scriptsize
    \begin{tabular}{l|ccccc|ccccc}
        & \multicolumn{5}{c|}{CD Tally} & \multicolumn{5}{c}{Lamp Return} \\
        Country & Gender & Age & Education & IncPctRank & Eskill & Gender & Age & Education & IncPctRank & Eskill \\ 
        \hline
        Austria & \textcolor{red}{-0.179} & \textcolor{red}{-0.080} & 0.120 & 0.010 & \textcolor{blue}{0.145} & \textcolor{blue}{0.382} & \textcolor{red}{-0.045} & \textcolor{blue}{0.263} & \textcolor{blue}{0.113} & 0.000 \\ 
        Belgium & -0.070 & \textcolor{red}{-0.076} & 0.030 & 0.050 & 0.060 & \textcolor{blue}{0.357} & -0.030 & \textcolor{blue}{0.363} & \textcolor{blue}{0.083} & -0.070 \\ 
        Germany & -0.100 & \textcolor{red}{-0.108} & 0.060 & 0.020 & \textcolor{blue}{0.217} & \textcolor{blue}{0.345} & \textcolor{red}{-0.077} & \textcolor{blue}{0.322} & \textcolor{blue}{0.096} & 0.010 \\ 
        Denmark & \textcolor{red}{-0.176} & \textcolor{red}{-0.093} & 0.020 & 0.020 & \textcolor{blue}{0.117} & \textcolor{blue}{0.368} & \textcolor{red}{-0.063} & \textcolor{blue}{0.195} & \textcolor{blue}{0.111} & -0.010 \\ 
        Estonia & -0.090 & \textcolor{red}{-0.090} & 0.040 & 0.000 & \textcolor{blue}{0.119} & \textcolor{blue}{0.261} & \textcolor{red}{-0.059} & \textcolor{blue}{0.262} & \textcolor{blue}{0.115} & -0.040 \\ 
        Finland & -0.110 & \textcolor{red}{-0.096} & 0.070 & 0.010 & \textcolor{blue}{0.145} & \textcolor{blue}{0.294} & \textcolor{red}{-0.067} & \textcolor{blue}{0.338} & \textcolor{blue}{0.064} & -0.010 \\ 
        United Kingdom & -0.010 & \textcolor{red}{-0.089} & -0.030 & 0.040 & \textcolor{blue}{0.138} & \textcolor{blue}{0.230} & \textcolor{red}{-0.038} & \textcolor{blue}{0.261} & \textcolor{blue}{0.054} & -0.010 \\ 
        Ireland & -0.020 & \textcolor{red}{-0.064} & 0.060 & 0.000 & \textcolor{blue}{0.178} & \textcolor{blue}{0.302} & -0.020 & \textcolor{blue}{0.353} & 0.030 & 0.000 \\ 
        South Korea & -0.010 & \textcolor{red}{-0.108} & 0.120 & 0.050 & 0.090 & \textcolor{blue}{0.317} & \textcolor{red}{-0.050} & \textcolor{blue}{0.323} & \textcolor{blue}{0.065} & -0.050 \\ 
        Netherlands & -0.070 & \textcolor{red}{-0.072} & 0.050 & 0.000 & \textcolor{blue}{0.110} & \textcolor{blue}{0.450} & \textcolor{red}{-0.049} & \textcolor{blue}{0.233} & \textcolor{blue}{0.131} & -0.060 \\ 
        Norway & -0.110 & \textcolor{red}{-0.078} & -0.030 & 0.020 & \textcolor{blue}{0.127} & \textcolor{blue}{0.330} & \textcolor{red}{-0.059} & \textcolor{blue}{0.283} & \textcolor{blue}{0.098} & -0.030 \\ 
        Poland & -0.090 & \textcolor{red}{-0.075} & 0.070 & 0.040 & 0.070 & \textcolor{blue}{0.239} & \textcolor{red}{-0.066} & \textcolor{blue}{0.400} & \textcolor{blue}{0.106} & -0.090 \\ 
        Slovakia & 0.120 & -0.050 & 0.010 & 0.080 & 0.090 & \textcolor{blue}{0.268} & -0.020 & \textcolor{blue}{0.400} & \textcolor{blue}{0.155} & -0.050 \\ 
        United States & -0.070 & \textcolor{red}{-0.077} & 0.080 & 0.030 & 0.100 & \textcolor{blue}{0.356} & \textcolor{red}{-0.061} & \textcolor{blue}{0.448} & \textcolor{blue}{0.110} & \textcolor{red}{-0.112} \\
    \end{tabular}
}
\caption{Posterior means of $\boldsymbol{\alpha}$ for the CD Tally and Lamp Return test items. Blue and red text colors represent significant positive and negative values, respectively.
}
\label{tab:beta-cov}
\end{table}

\paragraph*{Parameters $\beta_1$ and $\beta_2$}
Parameters $\beta_{c_i, 1}$ and $\beta_{c_i, 2}$ represent the impact of key actions on transition speed within the hazard function for group $c_i$. For group $c_i$, a larger $\beta_{c_i, 1}$ suggests a faster transition from the key action, while a larger $\beta_{c_i, 2}$ indicates a quicker transition to the key action. Statistical significance is determined using the 95\% Highest Posterior Density (HPD) interval, with parameters whose intervals include zero considered statistically insignificant.

\begin{table}[htbp]
\centering
{\small \renewcommand{\arraystretch}{1}
    \begin{tabular}{l|cc|cc|cc|cc}
        & \multicolumn{4}{c|}{CD Tally} & \multicolumn{4}{c}{Lamp Return} \\
        & \multicolumn{2}{c|}{Correct Group} & \multicolumn{2}{c|}{Incorrect Group} & \multicolumn{2}{c|}{Correct Group} & \multicolumn{2}{c}{Incorrect Group} \\
        Country & $\beta_{1,1}$ & $\beta_{1,2}$ & $\beta_{0,1}$ &$\beta_{0,2}$ & $\beta_{1,1}$ & $\beta_{1,2}$ & $\beta_{0,1}$ &$\beta_{0,2}$ \\ \hline
        Austria  & -0.270 & \textcolor{blue}{1.398} & -0.020 & 0.540 & \textcolor{red}{-1.996} & \textcolor{blue}{1.141} & \textcolor{red}{-1.954} & \textcolor{blue}{0.793} \\ 
        Belgium & -0.130 & \textcolor{blue}{1.822} & 0.410 & 0.550 & \textcolor{red}{-2.358} & \textcolor{blue}{1.244} & \textcolor{red}{-2.222} & \textcolor{blue}{0.928} \\ 
        Germany & -0.160 & \textcolor{blue}{1.655} & 0.310 & 0.670 & \textcolor{red}{-2.094} & \textcolor{blue}{1.055} & \textcolor{red}{-2.147} & \textcolor{blue}{1.007} \\ 
        Denmark & 0.060 & \textcolor{blue}{1.565} & 0.190 & \textcolor{blue}{0.949} & \textcolor{red}{-2.064} & \textcolor{blue}{1.239} & \textcolor{red}{-2.116} & \textcolor{blue}{1.240} \\ 
        Estonia & -0.280 & \textcolor{blue}{1.415} & 0.020 & \textcolor{blue}{0.943} & \textcolor{red}{-2.214} & \textcolor{blue}{1.226} & \textcolor{red}{-2.289} & \textcolor{blue}{1.029} \\ 
        Finland & 0.220 & \textcolor{blue}{1.162} & \textcolor{blue}{0.694} & 0.430 & \textcolor{red}{-2.138} & \textcolor{blue}{1.070} & \textcolor{red}{-2.176} & \textcolor{blue}{1.010} \\ 
        United Kingdom & 0.200 & \textcolor{blue}{1.353} & \textcolor{blue}{0.773} & 0.130 & \textcolor{red}{-1.622} & \textcolor{blue}{1.222} & \textcolor{red}{-1.579} & \textcolor{blue}{1.232} \\ 
        Ireland & 0.460 & \textcolor{blue}{0.927} & 0.520 & 0.470 & \textcolor{red}{-2.458} & \textcolor{blue}{1.418} & \textcolor{red}{-2.164} & \textcolor{blue}{0.871} \\ 
        South Korea & -0.330 & \textcolor{blue}{1.577} & -0.230 & \textcolor{blue}{0.959} & \textcolor{red}{-1.740} & \textcolor{blue}{0.570} & \textcolor{red}{-1.733} & \textcolor{blue}{0.215} \\ 
        Netherlands & 0.050 & \textcolor{blue}{1.477} & 0.370 & 0.460 & \textcolor{red}{-2.196} & \textcolor{blue}{1.104} & \textcolor{red}{-2.149} & \textcolor{blue}{0.918} \\ 
        Norway & 0.280 & \textcolor{blue}{1.452} & \textcolor{blue}{1.386} & 0.180 & \textcolor{red}{-2.416} & \textcolor{blue}{1.282} & \textcolor{red}{-2.381} & \textcolor{blue}{1.200} \\ 
        Poland & \textcolor{red}{-0.526} & \textcolor{blue}{1.508} & -0.49 & \textcolor{blue}{0.654} & \textcolor{red}{-2.222} & \textcolor{blue}{1.103} & \textcolor{red}{-2.194} & \textcolor{blue}{0.881} \\ 
        Slovakia & \textcolor{blue}{1.652} & -0.190 & \textcolor{blue}{1.687} & -0.270 & \textcolor{red}{-1.651} & 0.260 & \textcolor{red}{-1.584} & -0.130 \\ 
        United States & -0.110 & \textcolor{blue}{1.337} & 0.100 & \textcolor{blue}{0.790} & \textcolor{red}{-2.171} & \textcolor{blue}{0.813} & \textcolor{red}{-2.107} & \textcolor{blue}{0.544} \\ 
    \end{tabular}
}
\caption{Posterior means of key action effect ($\beta_{c_i, 1}$ and $\beta_{c_i, 2}$) for CD Tally and Lamp Return test items. Significant positive effects in blue, negative in red.}
\label{tab:beta_u03a}
\end{table}

In Table \ref{tab:beta_u03a} presents the estimated $\beta_{c_i, 1}$ and $\beta_{c_i, 2}$ values for 14 countries, comparing results for the CD Tally and Lamp Return test items. Significant positive and negative effects are marked in blue and red, respectively. 
\textcolor{black}{
For the CD Tally test item, both the correct and incorrect groups tend to exhibit statistically insignificant $\beta_{0,1}$ and $\beta_{1,1}$ values across most countries, suggesting that transitions from key actions generally have negligible effects.
} 

The analysis of $\beta_{c_i, 2}$, which represents the speed of transition to key actions, reveals an intriguing pattern. While the correct group demonstrates significant \textcolor{black}{positively} $\beta_{1, 2}$ values most countries, the incorrect group shows significant $\beta_{0, 2}$ in fewer countries. 
\textcolor{black}{
Notably, the magnitude of $\beta_{1, 2}$ generally exceeds that of $\beta_{0, 2}$, suggesting that the correct group transitions to key actions more rapidly than the incorrect group.
}

\subsubsection{Compare Transition Speed Between Two Groups}

To compare transition patterns between correct and incorrect answer groups, we analyzed differences in transition probabilities using parameters estimated from our proposed model. Transition probabilities were calculated using Equation \ref{eq:prob}, with covariates set to specific values: Gender = 1, Age = 5, Education = 2, IncPctRank = 4, and Eskill = 1. Statistical significance of differences was determined using 95\% HPD intervals. Figure \ref{fig:diff-cdtally} illustrates these differences for the USA using  two visualization methods: a heatmap (Figure \ref{fig:diff-a-cdtally}) and a network diagram (Figure \ref{fig:diff-b-cdtally}). These visualizations clearly show notable differences in action transition speeds between the two groups.

\begin{figure}[htbp]
    \centering
    \begin{subfigure}[b]{0.45\textwidth}
         \centering
         \includegraphics[width= \textwidth, page = 1]{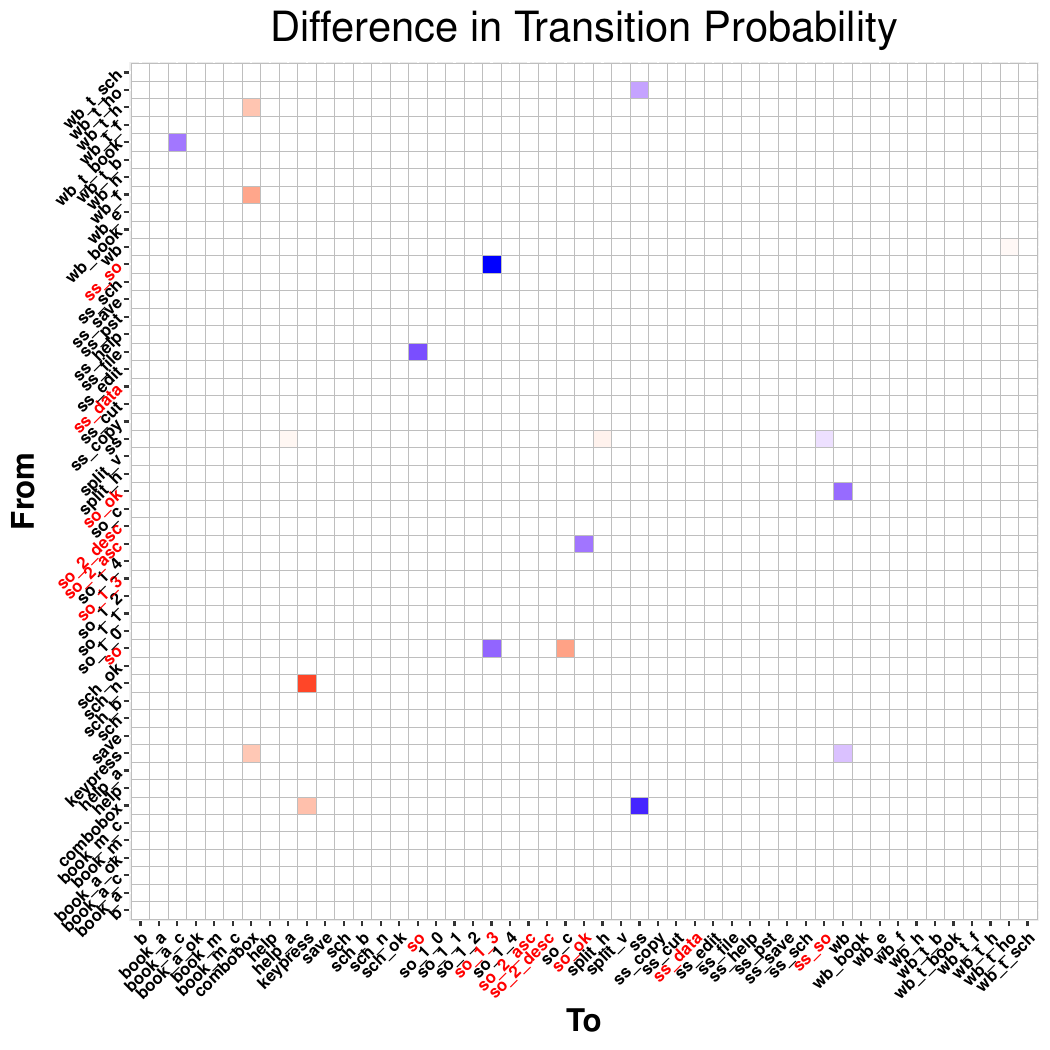}  
         \caption{Heatmap}
         \label{fig:diff-a-cdtally}
     \end{subfigure}
     \begin{subfigure}[b]{0.45\textwidth}
         \centering
         \includegraphics[width= \textwidth, page = 2]{Figure/transition_diff_US.pdf}  
         \caption{Network visualization}
         \label{fig:diff-b-cdtally}
     \end{subfigure} 
    \caption{
    Transition probability differences between correct and incorrect groups for USA CD Tally test item: (a) heatmap and (b) network visualization. Blue indicates higher probability for correct group, red for incorrect. Color intensity in (a) and the arrow thickness in (b) represents difference magnitude between two groups. Key actions are highlighted in red text.}
    \label{fig:diff-cdtally}
\end{figure}

Figure \ref{fig:diff-a-cdtally} uses a color-coded heatmap to illustrate transition probability differences between correct and incorrect groups. \textcolor{black}{X and Y axes represent the ``to'' and ``from'' action state}, respectively, where key actions are colored in red. Blue cells represent transitions where the correct answer group has significantly higher probabilities, red cells indicate the incorrect group showing significantly higher probabilities.  White cells represent non-significant differences between the two groups. The color intensity corresponds to the magnitude of these differences. Notable patterns emerge, such as the blue cell for the `combobox' (selecting a combobox value) to `ss' (switching to the spreadsheet page) transition, indicating faster movement from `combobox' to `ss' by correct respondents. Conversely, the red cells like the `sch\_n' (clicking the next button in the search engine) to `keypress' (pressing a keyboard key) transition suggest quicker progression by the incorrect group in certain actions.

Figure \ref{fig:diff-b-cdtally} presents a network diagram of actions (nodes) and transitions (directed edges), where key actions are colored in red text. To improve the clarity of the presentation, we displayed only statistically significant differences in the graph. Arrow thickness represents the magnitude of the probability differences, with blue arrows indicating faster transitions by the correct group and red arrows indicating faster transitions by the incorrect group. This visualization reveals that the correct answer group typically executes transitions that involve key actions, which are crucial to reaching the correct answer, more rapidly. Key examples include faster transitions from `so\_2\_asc' (sorting the spreadsheet in ascending order) to `so\_ok' (clicking `Ok' after setting sorting options), `ss\_so' (clicking the sort engine on the spreadsheet page) to `so\_1\_3' (sorting by the third column, Genre), and `so' (clicking the sort engine through the data menu on the spreadsheet page) to `so\_1\_3' (sorting by the third column, Genre). These transactions that occur when sorting the spreadsheet to solve the problem are critical parts of the process of solving the correct answer. However, the incorrect group showed faster execution of actions that might be less directly related to solving the problem. For instance, `wb\_t\_h' (clicking the help button on the website page toolbar) to `combobox' (selecting a combobox value) and `wb\_f' (clicking the file menu on the website page) to `combobox' were performed more rapidly by the incorrect group.

\subsection{Lamp Return}

\subsubsection{Parameter Estimation}
\paragraph*{Parameter $\kappa_{c_i,m,l}$}

Table \ref{tab:kappa_ml_lamp} shows the five highest and lowest baseline hazard parameters ($\kappa_{c_i,m,l}$) for the USA in the Lamp Return test item, indicating the transition probabilities from action $m$ to $l$ for correct ($\kappa_{1,m,l}$) and incorrect ($\kappa_{0,m,l}$) groups. The results for other countries provided in the Section 4 of the Supplementary Material. Higher $\kappa_{c_i,m,l}$ values indicate a greater transition likelihood from action $m$ to $l$ for group $c_i$.

For both correct and incorrect groups, the top five transitions are similar, primarily involving web page navigation. A notable example is the transition from \textcolor{black}{`wb\_pg\_8\_3' (Link to obtain authorization number on the Customer Service page) to `wb\_pg\_8\_3\_1' (Request authorization number on the Customer Service page), which exhibits high $\kappa$ values of 20.54 and 20.15 for incorrect and correct groups, respectively. } In contrast, the transitions from `wb\_pg\_8\_3\_1' and `wb\_pg\_pop2' (Click the close button on pop-up system message 2) have low $\kappa$ values for both groups. In conclusion, for the Lamp Return test item, the baseline hazard was similar between the correct and incorrect answer groups.

\begin{table}[htbp]
\centering
{\small
\textcolor{black}{
    \begin{tabular}{c|llr|llr}
        & \multicolumn{3}{c|}{Top 5} & \multicolumn{3}{c}{Bottom 5}\\
        & From & To & Value & From & To & Value \\ 
          \hline
        \multirow{5}{*}{Incorrect ($\kappa_{0, m, l}$)}  & wb\_pg\_8\_3 & wb\_pg\_8\_3\_1 & 20.54 & wb\_pg\_8\_3\_1 & wb & 0.01 \\ 
        & wb\_pg\_8\_2 & wb\_pg\_8 & 15.34 & wb\_pg\_8\_3\_1 & wb\_hist\_back & 0.01 \\ 
        & paste & keypress & 11.86 & wb\_pg\_pop2 & file & 0.02 \\ 
        & wb\_pg\_8\_4\_submit & wb\_pg\_pop3 & 11.22 & wb\_pg\_pop2 & wb\_pg\_0 & 0.02 \\ 
        & wb\_pg\_8\_1 & wb\_pg\_8 & 11.07 & wb\_pg\_pop1 & wb\_pg\_1\_2 & 0.02 \\ \hline
        \multirow{5}{*}{Correct ($\kappa_{1, m, l}$)} & wb\_pg\_8\_3 & wb\_pg\_8\_3\_1 & 20.15 & wb\_pg\_8\_3\_1 & wb & 0.01 \\ 
        & wb\_pg\_8\_2 & wb\_pg\_8 & 18.19 & wb\_pg\_8\_3\_1 & wb\_hist\_back & 0.01 \\ 
        & wb\_pg\_8\_4\_submit & wb\_pg\_pop3 & 9.64 & wb\_pg\_2 & wb & 0.02 \\ 
        & wb\_pg\_8\_1 & wb\_pg\_8 & 9.43 & wb\_pg\_pop2 & file & 0.02 \\ 
        & wb\_pg\_8\_4\_reason\_4 & wb\_pg\_8\_4\_request\_1 & 5.83 & wb\_pg\_pop2 & wb\_hist\_for & 0.02 \\  
    \end{tabular}
    }}
\caption{
The five highest and lowest $\kappa_{1, m, l}$ and $\kappa_{0, m, l}$ for correct and incorrect groups in the USA Lamp Return test item.}
\label{tab:kappa_ml_lamp}
\end{table}

\paragraph*{Parameter $\tau$}

The parameter $\tau_i$ quantifies the overall action transition speed for each respondent $i$, with higher values indicating faster transitions. Figure \ref{fig:tau-lamp} represents boxplots of the estimated $\tau_i$ across countries for the Lamp Return test item. 
\textcolor{black}{
Similar to the CD Tally test item, boxplots are also approximately symmetric and centered around 1 across all countries under baseline characteristics (i.e., when all covariates are set to zero).
}

\begin{figure}[htbp]
    \centering
    \includegraphics[width = .4\textwidth]{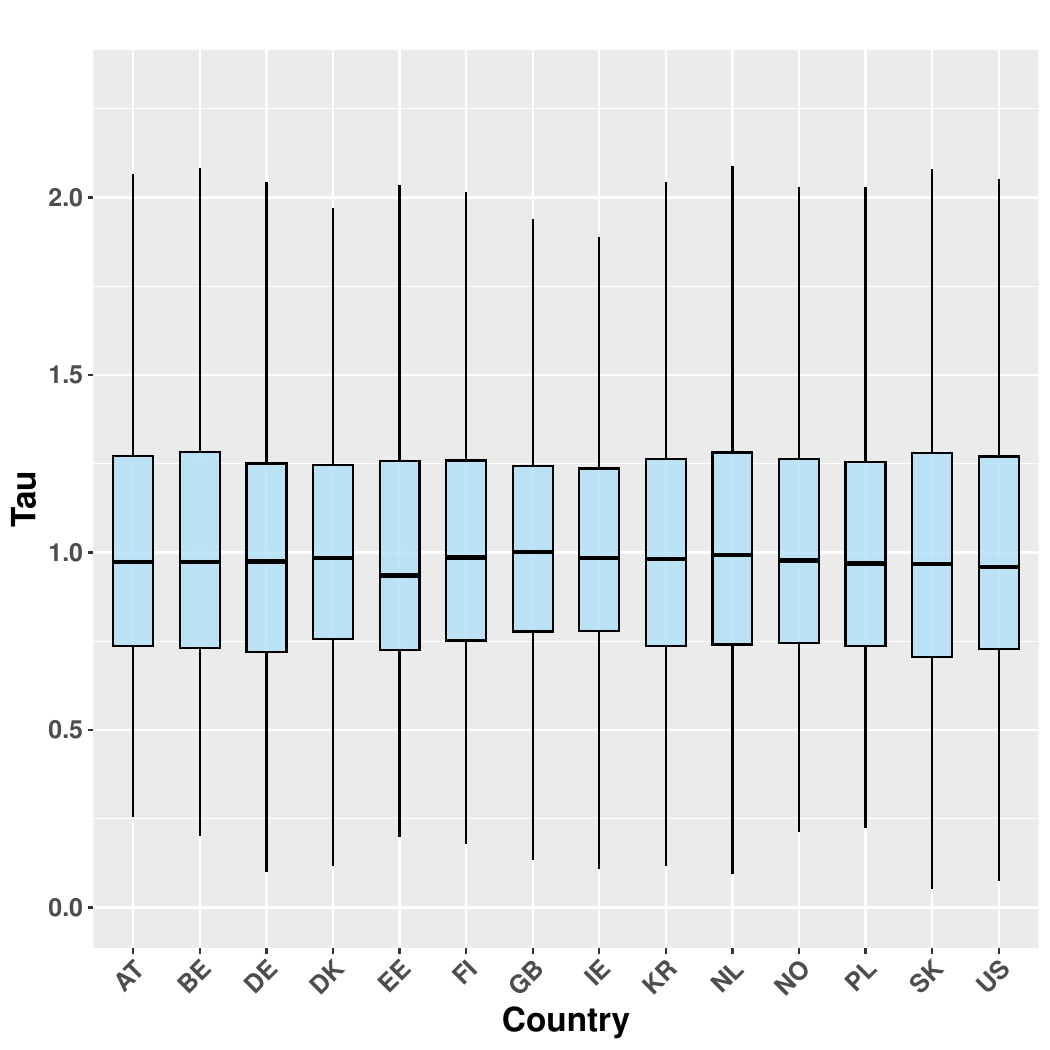}
    \caption{\textcolor{black}{The boxplots of posterior means of individual transition speed parameters ($\tau_i$) across the 14 countries for Lamp Return test item.} Outliers omitted.}
    \label{fig:tau-lamp}
\end{figure}

\paragraph*{Parameters $\boldsymbol{\alpha}$}
The parameters $\alpha$ represent the effect of individual covariates on the transition speed.
Table \ref{tab:beta-cov} presents the estimated $\boldsymbol{\alpha}$ values for the Lamp Return test item, where a color-coding scheme is applied to highlight statistical significance. The blue entries denote significantly positive effects, while the red entries indicate significantly negative effects. 
The analysis of covariate effects reveals consistent patterns across most countries for the Lamp Return test item. Gender, education level, and income level demonstrate positive significance, while age shows negative significance. These findings suggest that female respondents, younger test-takers, and individuals with higher education and income levels tend to exhibit faster action transition speeds. 

\paragraph*{Parameters $\beta_1$ and $\beta_2$}

Table \ref{tab:beta_u03a} presents the estimated values of $\beta_{c_i, 1}$ and $\beta_{c_i, 2}$ for the Lamp Return test item across 14 countries, with these parameters quantifying the impact of key actions on transition speeds. A larger $\beta_{c_i, 1}$ indicates faster transitions from key actions, while a larger $\beta_{c_i, 2}$ signifies quicker transitions to key actions for group $c_i$. In Table \ref{tab:beta_u03a} color-coding  is used to highlight statistical significance, with blue denoting significant positive effects and red indicating significant negative effects.

Table \ref{tab:beta_u03a} for the Lamp Return test item reveals that there are consistent patterns across countries in terms of the impact of key actions on transition speeds. Both $\beta_{0, 1}$ and $\beta_{1, 1}$ show negative significance universally, indicating slower transitions from key actions, while $\beta_{0, 2}$ and $\beta_{1, 2}$ display positive significance in most countries, suggesting faster transitions to key actions for both groups.  
\textcolor{black}{
Notably, in many countries, the absolute values of $\beta_{1,1}$ and $\beta_{1,2}$ slightly exceed those of $\beta_{0,1}$ and $\beta_{0,2}$, respectively. This pattern suggests that the correct group generally moves away from key actions more slowly but approaches them more quickly than the incorrect group.
}

\subsubsection{Compare Transition Speed Between Two Groups}

To compare transition patterns between correct and incorrect groups for the Lamp Return test item using estimated parameters and applied a 95\% HPD interval to assess the statistical significance of transition probability differences.  Figure \ref{fig:diff-lamp} illustrates these differences using a network diagram based on specific covariate values (Gender = 1, Age = 5, Education = 2, IncPctRank = 4, Eskill = 1).  Only significant differences in transition probabilities between the two groups are shown in Figure \ref{fig:diff-lamp}. Due to the large number of actions in Lamp Return test item, we focus on the network visualization results here. 

\begin{figure}[htbp]
    \centering
    \begin{subfigure}[b]{0.45\textwidth}
         \centering
         \includegraphics[width= \textwidth, page = 2]{Figure/transition_diff_US_sep.pdf}  
         \caption{}
         \label{fig:diff-a-lamp}
     \end{subfigure}
     \begin{subfigure}[b]{0.45\textwidth}
         \centering
         \includegraphics[width= \textwidth, page = 1]{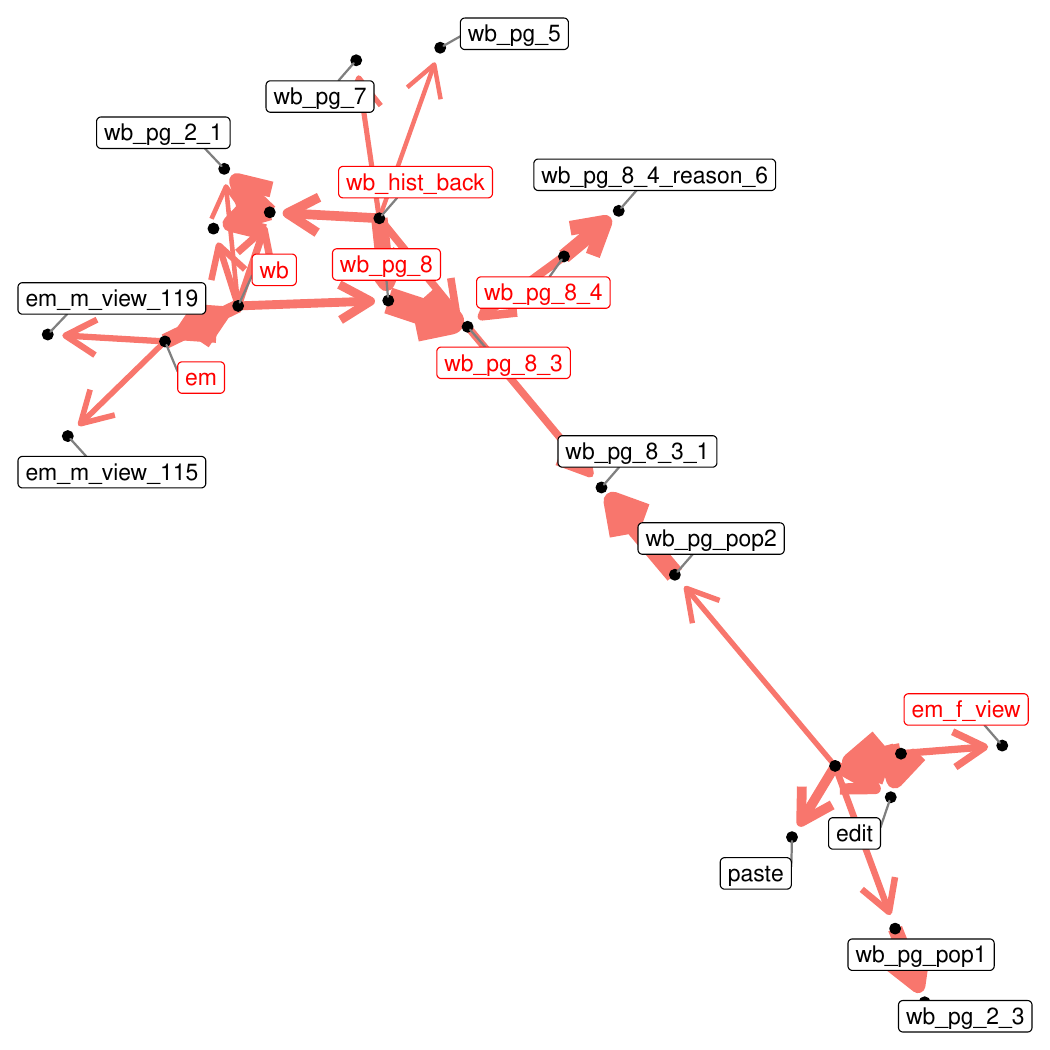}  
         \caption{}
         \label{fig:diff-b-lamp}
     \end{subfigure} 
    \caption{
        Network visualization of transition probability differences for the Lamp Return test item in the USA. (a) shows significantly higher transition probabilities for the correct group (blue arrows), while (b) shows higher probabilities for the incorrect group (red arrows). The arrow thickness indicates the magnitude of the difference in transition probability between the two groups. Key actions are highlighted with red text.
        }
    \label{fig:diff-lamp}
\end{figure}

Figures \ref{fig:diff-a-lamp} and \ref{fig:diff-b-lamp} illustrate significant differences in transition probabilities between correct and incorrect groups for the Lamp Return task, respectively. Figure \ref{fig:diff-a-lamp} reveals that the correct group exhibited faster transitions crucial to solving the problem, such as  `wb\_pg\_8' (Link to Customer Service page) to `wb\_pg\_8\_4' (Link to view return form on the Customer Service page) and  `wb\_pg\_8\_4' to `wb\_pg\_8\_4\_reason\_4' (Select the second reason for return (Wrong item shipped)),  `wb\_pg\_8\_4\_request\_1' (Select a first request for returned items (Exchange for the correct item)) to `em' (Switch to Email page), and `em' to `em\_m\_view\_305' (View email 305 (confirm authorization number) on the email page). These transitions represent key steps in the return process, including finding the authorization number via email, selecting a reason to return an item, and submitting a return request. Conversely, Figure \ref{fig:diff-b-lamp} shows that the incorrect group had faster transitions primarily related to general website navigation, such as  `wb\_pg\_2\_1' (Link to a first sub-page on the Desk Lamps page), `wb\_pg\_5' (Link to the New Arrivals page on the web page), `wb\_pg\_7' (Link to Customer Comments page on the web page),  and `wb\_pg\_8' (Link to Customer Service page). This contrast suggests that while the correct group more efficiently executed task-critical actions, the incorrect group spent more time on rapid but less targeted website exploration.

The analysis of transition patterns reveals a clear distinction between successful and unsuccessful problem-solving strategies in the Lamp Return task. Respondents who provided correct answers demonstrated faster transitions through task-critical actions directly related to the solution, while those who answered incorrectly exhibited quicker transitions in general website navigation but slower progression through solution-specific steps.

{\color{black}
\section{Model Evaluation and Robustness}
\subsection{Sensitivity Analysis}
In Bayesian analysis, prior distributions can have a significant impact on posterior inferences, especially when data is limited or the model structure is complex. Consequently, it is crucial to assess how sensitive the results are to different prior specifications \citep{Depaoli2020}. Conducting a prior sensitivity analysis enables us to test the robustness of our findings by comparing posterior estimates across various prior distributions. This process ensures that the results are not unduly influenced by particular prior choices, thus improving the reliability of the conclusions.
 
We performed a prior sensitivity analysis on a set of key model parameters ${\boldsymbol{\kappa}, \boldsymbol{\tau}, \boldsymbol{\alpha}, \boldsymbol{\beta}}$ to evaluate the robustness of the posterior estimates under different levels of prior informativeness. To investigate the effect of prior informativeness, we created a range of specifications, from highly informative to weak or non-informative priors, by adjusting the hyperparameters. 
For $\boldsymbol{\beta}$ and $\boldsymbol{\alpha}$, we set normal priors with standard deviations $\sigma_\beta$ and $\sigma_\alpha$ set to 0.5, 1, 2, 5, and 10. For $\boldsymbol{\kappa}$ and $\boldsymbol{\tau}$, we considered Gamma$(a, b)$ priors with the following shape and scale combinations: (1, 1), (0.5, 2), (0.1, 10), (0.01, 100), and (0.001, 1000).

Table \ref{tab:sens-beta} shows 
the posterior estimates for $\boldsymbol{\beta}$ across a range of prior standard deviations, $\sigma_\beta \in \{0.5, 1, 2, 5, 10\}$. The results suggest that increasing $\sigma_\beta$ from 0.5 to 2 leads to changes in posterior means, especially for $\beta_{1,2}$ and $\beta_{0,2}$. However, with $\sigma_\beta \geq 2$, the posterior estimates remain  relatively stable for all $\boldsymbol{\beta}$ parameters, with minimal changes in both central tendency and dispersion. These findings suggest that the inferences on $\boldsymbol{\beta}$ can  be robust to prior specification with  $\sigma_\beta = 2$. 
Accordingly, we selected $\sigma_\beta = 2$ for the main analyses to balance prior flexibility with inferential stability. Trace plots for each prior setting, available in the project’s GitHub repository, further confirm consistent convergence and reliable posterior recovery.

The results for the remaining parameters, detailed in Section 7 of the Supplementary Material, demonstrate that posterior means, standard deviations, and 95\% HPD intervals remained largely stable across different prior hyperparameters, indicating robustness to reasonable variations in prior settings.

\begin{table}[h]
\centering
{\scriptsize \color{black}
\begin{tabular}{ccc|ccc}
Group & Parameter & $\sigma_{\beta}$ & Mean & SD & HPD Interval \\ 
  \hline
\multirow{10}{*}{Correct} & \multirow{5}{*}{$\beta_{1, 1}$} &
0.5 & 0.11 & 0.26 & $[-0.391,\ 0.62]$ \\
  & & 1 & -0.01 & 0.29 & $[-0.582,\ 0.544]$ \\
  & & 2 & -0.11 & 0.31 & $[-0.704,\ 0.501]$ \\
  & & 5 & -0.11 & 0.30 & $[-0.688,\ 0.481]$ \\
  & & 10 & -0.12 & 0.29 & $[-0.673,\ 0.476]$ \\ \cline{2-6}
  & \multirow{5}{*}{$\beta_{1, 2}$} &
  0.5 & 1.00 & 0.25 & $[0.519,\ 1.506]$ \\
  & & 1 & 1.22 & 0.28 & $[0.668,\ 1.76]$ \\
  & & 2 & 1.34 & 0.32 & $[0.72,\ 1.955]$ \\
  & & 5 & 1.34 & 0.30 & $[0.746,\ 1.902]$ \\
  & & 10 & 1.35 & 0.30 & $[0.787,\ 1.98]$ \\
  \hline
\multirow{10}{*}{Incorrect} & \multirow{5}{*}{$\beta_{0, 1}$} &
0.5 & 0.16 & 0.26 & $[-0.356,\ 0.67]$ \\
  & & 1 & 0.12 & 0.31 & $[-0.499,\ 0.707]$ \\
  & & 2 & 0.10 & 0.32 & $[-0.527,\ 0.711]$ \\
  & & 5 & 0.09 & 0.32 & $[-0.534,\ 0.728]$ \\
  & & 10 & 0.10 & 0.32 & $[-0.53,\ 0.732]$ \\ \cline{2-6}
  & \multirow{5}{*}{$\beta_{0, 2}$} &
0.5 & 0.60 & 0.26 & $[0.078,\ 1.097]$ \\
  & & 1 & 0.75 & 0.31 & $[0.168,\ 1.37]$ \\
  & & 2 & 0.79 & 0.32 & $[0.155,\ 1.426]$ \\
  & & 5 & 0.81 & 0.33 & $[0.136,\ 1.417]$ \\
  & & 10 & 0.81 & 0.33 & $[0.156,\ 1.448]$ \\
\end{tabular}
}
\caption{
Results of prior sensitivity analysis of $\beta_{1, 1}$, $\beta_{1, 2}$, $\beta_{0, 1}$, and $\beta_{0, 2}$ for USA CD Tally test item. The table summarizes the posterior means, standard deviations, and 95\% HPD intervals under different prior standard deviations, $\sigma_{\beta}$.
}
\label{tab:sens-beta}
\end{table}
}

{\color{black}
\subsection{Simulation Study}

Next, we conducted a simulation study to evaluate the estimation accuracy of our proposed MSM approach. We designed four simulation scenarios, each representing a different level of group heterogeneity and key action effects. The scenarios are summarized in Table~\ref{tab:sim_scenarios}, highlighting differences in covariate effects, key action effects, and the presence or absence of group-level distinctions.

\begin{table}[H]
\centering
{\scriptsize \color{black}
\begin{tabular}{c|c|c|c|c}
\textbf{Scenario} & \textbf{Description} & \textbf{Covariate Effects} & \textbf{Key Action Effects} & \textbf{Group Differences} \\
\hline
1 & Baseline (No Group Differences) & Homogeneous & Homogeneous & None \\
2 & Heterogeneous Covariate Effects & Heterogeneous & Homogeneous & None \\
3 & Group Differences (Start and End) & Homogeneous & Distinct by group (start and end) & Yes \\
4 & Group Differences (One Side Only) & Homogeneous & Distinct by group (one side only) & Yes \\
\end{tabular}
}
\caption{Summary of simulation scenarios.}
\label{tab:sim_scenarios}
\end{table}

For all scenarios, we generated data for 400 individuals, each performing a sequence of actions selected from a set of 50 possible actions, which included 10 predefined key actions. We assumed that all transitions between actions were possible, as physical or interface constraints limiting transitions between actions were not available in the simulation. While this simplification departs somewhat from real-world settings, it still allows us to assess the model's statistical properties in a controlled simulation environment. We included five covariates, each independently drawn from a normal distribution with a mean of 0 and a variance of 4. The number of actions per individual was sampled from a negative binomial distribution with parameters $r = 5$ and $p = 1/3$, selected to approximate the over-dispersion in action counts observed in real log data. Further details of the simulation scenarios are provided in Section 8 of the Supplementary Material.

To evaluate estimation performance across the simulation scenarios, we compared the estimated transition probabilities with the true probabilities used in data generation. For each simulation run:
\begin{enumerate}
    \item We computed the MSE for each individual by comparing their estimated and true $50 \times 50$ transition probability matrices, averaging the squared differences over all 2,500 entries.
    \item We then averaged the individual-level MSEs to obtain a single summary measure for the simulation run.
    \item This procedure was repeated 200 times per scenario.
\end{enumerate}
The resulting distributions of simulation-level MSEs were visualized using boxplots to assess estimation accuracy. Lower MSE values indicate more precise recovery of the true transition structure.

As shown in Figure \ref{fig:simul}, the model demonstrates strong recovery performance across all four simulation scenarios. The mean estimation errors for Scenarios 1 through 4 were $3.5458\times 10^{-4}$, $3.5458 \times 10^{-4}$, $4.9466 \times 10^{-4}$, and $3.7993 \times 10^{-4}$, respectively. Scenarios 3 and 4, which include varying levels of group heterogeneity, exhibited slightly higher errors than Scenarios 1 and 2. However, estimation errors were minimal and tightly concentrated around zero across all conditions, indicating excellent parameter recovery. These results suggest that the proposed MSM approach is robust to increasing model complexity and effectively recovers transition structures under a range of realistic conditions.

\begin{figure}[ht]
    \centering
    \includegraphics[width=0.5\linewidth]{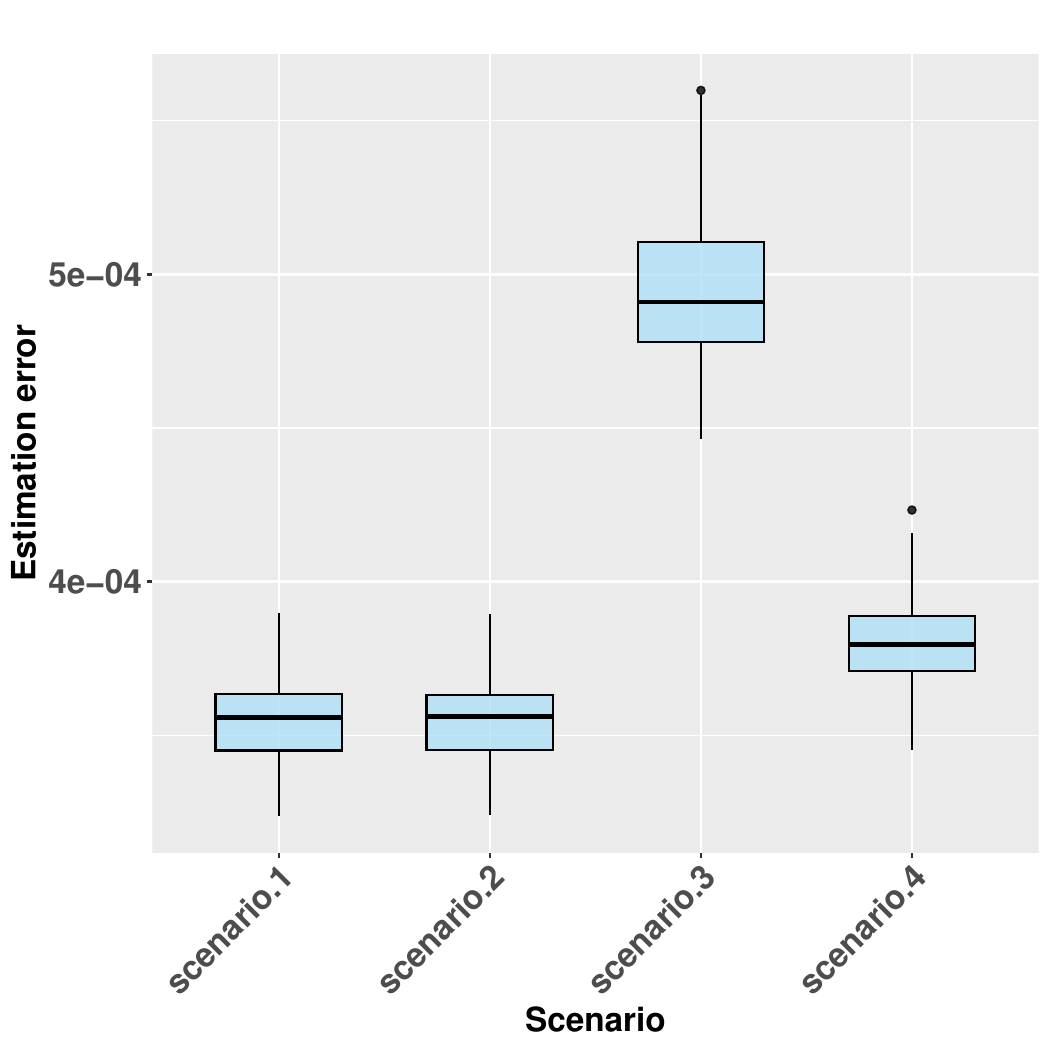}
    \caption{
    Boxplots summarizing the estimation errors across 200 simulation replications for each scenario. Estimation error is defined as the mean squared error (MSE), calculated by averaging the element-wise squared differences between the estimated and true transition probability matrices for each individual, and then averaging across individuals per run. Each boxplot reflects the distribution of these simulation-level MSEs under varying levels of group heterogeneity and covariate effects.
    }
    \label{fig:simul}
\end{figure}
}

\section{Conclusion}

\subsection{Summary}

The Program for the International Assessment of Adult Competencies (PIAAC), conducted by the OECD, evaluates adults' literacy, numeracy, and problem-solving skills in technology-rich environments (PSTRE). During the PSTRE assessment, user interactions with the computer, such as button clicks, links, dragging, dropping, copying, and pasting, are recorded in sequence with timestamps. Recorded sequence of events and actions as such constitute log data which is also called process data. Various methodologies have been developed to analyze log data. Studies have utilized timing data, such as the overall evaluation time of log data or event timestamps, for clustering or group comparisons.

Research on test-taking behavior has increasingly recognized the crucial role of time information, extending beyond total task duration to include transition times between actions in log data \citep{Goldhammer2014, Scherer2015, Vrs2016, Engelhardt2019, He2019b}. The analysis of transition time between actions can provide unique insights into respondents' cognitive processing, engagement levels, and problem-solving strategies. By examining the dynamics of action transitions, we can also learn about how individuals navigate complex tasks, identifies potential areas of difficulty, and illuminates the subtle differences in problem-solving approaches between correct and incorrect response groups \citep{Xu2020, Wang2022, Ulitzsch2021, Chen2020, Fu2023}.

In this study, we applied a multi-state survival (MSM) model for analyzing log data, by treating user actions as distinct states to capture the dynamic nature of problem-solving processes. To the best of our knowledge, this is a unique application of the model which, therefore, can be seen as an important contribution of our work. 

With MSM, we focus on examining how key actions and covariates influence transition speeds and patterns, particularly between correct and incorrect response groups. By employing the $\chi^2$ statistical approach to identify key actions \citep{He2015}, we differentiate between actions associated with correct and incorrect responses. This novel application of MSM to log data analysis which specific actions that significantly impact transition speeds and overall performance. 

We applied the proposed model to two problem-solving test items: CD Tally and Lamp Return. The baseline hazard parameters ($\boldsymbol{\kappa}$ and $\tau$) for correct and incorrect groups were similar across both items, indicating minimal differences in underlying transition likelihoods. Among 17,441 respondents, 4,233 (24.27\%) completed both items, with a weak but statistically significant correlation (0.1851, p-value $< 2.2 \times 10^{-16}$) between their $\tau$ values. 

Covariate effects, $\boldsymbol{\alpha}$, differed between two test items; for the CD Tally test item, younger age and higher computer skills (`Eskill') correlated with faster transition, while for the Lamp Return test item, all covariates except Eskill were significant. The impact of key actions, $\boldsymbol{\beta}$, also varied between the two test items; in the CD Tally test item, transitions were faster 
\textcolor{black}{
when both the end actions were key actions, for the correct group,
} 
while in the Lamp Return test item, transitions were quicker from non-key actions to key actions, independent of correctness. Notably, comparing the transition probabilities between correct and incorrect groups revealed that correct respondents were quicker in transitions that are closely related to the correct answer for both CD Tally and Lamp Return test items. 
\textcolor{black}{
The observed differences between the two items can be attributed to the fundamentally distinct nature of their task environments and cognitive demands. The CD Tally item presents a structured problem-solving task within a spreadsheet environment, where transitions between actions follow more systematic and efficiency-driven. In this constrained environment, the EsKill (Electronic Skill) emerges as a significant predictor of transition speed, suggesting that technical proficiency directly influences the efficiency of navigating through spreadsheet-based problem-solving steps. In contrast, the Lamp Return item involves navigating through a simulated e-commerce website with multiple pathways and potential solutions, creating a more exploratory task environment. Here, significant predictors shift to demographic variables like gender and income, which likely reflect broader differences in the experience of online shopping, information seeking preferences, and decision-making approaches in less structured environments. The diminished influence of EsKill in this context suggests that general digital literacy may be less predictive of performance than specific domain familiarity and decision-making tendencies. Regarding the differences in key action effects ($\beta_{c_i, \cdot}$), our analysis reveals item-specific patterns that warrant further investigation. These differences likely stem from a combination of interface design elements, cognitive processing requirements, and the nature of successful strategies unique to each task. 
}

\subsection{Outlook}

While this paper focuses on PIAAC log data, the proposed multi-state survival model (MSM) has broad applicability across various domains. It can be used to analyze students' learning processes in one educational teaching and learning platforms, patient interactions with digital health platforms, consumer behavior in e-commerce, and employee task navigation in workplace training scenarios. This versatility allows for the identification of critical decision-making points, engagement patterns, and skill development opportunities in each respective field. 

We would like to focus on several areas to further improve the proposed MSM for log data analysis in future. First, further investigation is needed to utilize network analysis techniques, which was used in visualizing transition probabilities in the current paper, as it could provide richer insights into behavioral differences between correct and incorrect answer groups. Second, developing a robust model fit evaluation method is crucial for ensuring model reliability and applicability across diverse datasets. Third, improving and standardizing log data pre-processing methods is essential, as variations in pre-processing can significantly impact results and interpretation of the results. Fourth, incorporating a hierarchical structure into the model would enable meaningful cross-country comparisons, offering insights into how problem-solving behaviors vary across different cultural contexts and provide more comprehensive guidance for educational and policy interventions on a global scale. \textcolor{black}{Finally, extending the sensitivity analysis beyond a single-country dataset and designing more realistic simulation scenarios would strengthen the generalizability of the findings and offer additional validation of the model's robustness under diverse empirical conditions.}

\textcolor{black}{
In conclusion, MSMs offer unique analytical capabilities that extend beyond traditional methods. By modeling both the sequence and timing of transitions between cognitive states, MSMs capture the dynamic structure of problem-solving processes, revealing that correct and incorrect respondents diverge not only in their outcomes but also in their underlying behavioral trajectories. Unlike standard sequence analysis, MSMs further quantify the duration spent in each state, enabling the identification of cognitive bottlenecks, hesitation patterns, and differences in strategic execution. In particular, our findings highlight specific transition points where incorrect respondents exhibit prolonged delays - patterns that would remain obscured in aggregated response times or accuracy metrics. Moreover, MSMs permit the inclusion of covariates to examine how learner characteristics influence not just performance outcomes but also the problem-solving process itself. Our results demonstrate that certain demographic variables predict distinct strategic behaviors, underscoring the potential of MSMs to inform more targeted and individualized educational interventions.
}  

\section*{Acknowledgement}
\subsection*{Funding}
We thank the editor, associate editor, and reviewers for their constructive comments. This work was supported by the National Research Foundation of Korea [grant number NRF-2020R1A2C1A01009881, NRF-2021S1A3A2A03088949, RS-2023-00217705, and RS-2024-00333701; Basic Science Research Program awarded to IHJ] and Yonsei University Research Graph of 2024 [awarded to IHJ]. Correspondence should be addressed to Ick Hoon Jin, Department of Applied Statistics, Department of Statistics and Data Science, Yonsei University, Seoul, Republic of Korea. E-Mail: ijin@yonsei.ac.kr. 

\subsection*{Data Availability Statements}
The datasets generated during and/or analyzed during the current study are available in the GESIS, \url{https://doi.org/10.4232/1.12955}

\newpage

\bibliographystyle{Chicago}
\bibliography{reference.bib}

\end{document}